\documentclass{ieeeaccess}
\usepackage{cite}
\usepackage{amsmath,amssymb,amsfonts}
\usepackage{algorithmic}
\usepackage{graphicx}
\usepackage{textcomp}

\usepackage[english]{babel}

\usepackage{setspace}
\usepackage{color}
\usepackage{xcolor}
\definecolor{tvgred}{rgb}{1.00,0.00,1.00}
\definecolor{tvgviolet}{rgb}{1.00,0.00,1.00}

\newcommand{\Z}[1]{Z_\mathcal{#1}} 
\newcommand{\zt}[1]{z^{\mathcal{#1}}_t} 
\renewcommand{\S}[1]{\ensuremath{\mathbf{#1}}} 

\newtheorem{definition}{Definition}

\def\BibTeX{{\rm B\kern-.05em{\sc i\kern-.025em b}\kern-.08em
    T\kern-.1667em\lower.7ex\hbox{E}\kern-.125emX}}
\begin{document}
\history{Submitted, date of current version September 1, 2022.}
\doi{10.1109/ACCESS.2017.DOI}

\title{Indirect Dynamic Negotiation in the Nash Demand Game}
\author{\uppercase{\uppercase{Tatiana V. Guy}\authorrefmark{1},
		\IEEEmembership{Senior Member, IEEE}, Jitka Homolov\'{a}\authorrefmark{2},  Aleksej Gaj\authorrefmark{3}.}}
\address[1]{Department of Information Engineering, Faculty of Economics and Management, Czech University of Life Sciences}
\address[2]{Department of Adaptive Systems, Institute of Information Theory and Automation, The Czech Academy of Sciences}
\address[3]{Department of Mathematics, Faculty of Nuclear Sciences and Physical Engineering, Czech Technical University}
\tfootnote{This work was supported in part by the project LTC18075}

\markboth
{Guy \headeretal: Indirect Dynamic Negotiation in the Nash Demand Game}
{Guy \headeretal: Indirect Dynamic Negotiation in the Nash Demand Game}

\corresp{Corresponding author: Tatiana V. Guy (e-mail: guy@ieee.org).}

\begin{abstract}
The paper addresses a problem of sequential bilateral bargaining with incomplete information. We proposed a decision model that helps agents to successfully bargain by performing indirect negotiation and learning the opponent's model. Methodologically the paper casts heuristically-motivated bargaining of a self-interested independent player into a framework of Bayesian learning and Markov decision processes. The special form of the reward implicitly motivates the players to negotiate indirectly, via closed-loop interaction. We illustrate the approach by applying our model to the Nash demand game, which is an abstract model of bargaining. The results indicate that the established negotiation: i) leads to coordinating players' actions; ii) results in maximising success rate of the game and iii) brings more individual profit to the players.
\end{abstract}

\begin{keywords}
Learning, Markov decision process, Nash demand game, Negotiation.
\end{keywords}

\titlepgskip=-15pt

\maketitle

\section{Introduction}
\label{sec:Introduction}
\PARstart{P}{olitics} and business are considered traditional spheres of human negotiation. The internet and modern means of communication have extended human negotiation to new domains such as social networks, deliberative democracy, e-commerce, cloud-based applications, \cite{YanXie:20}, \cite{DavGerSarShe:10}. Besides, automatic bargaining and negotiation, being inevitable in modern cyber-physical-social systems \cite{LiLaiZhu:19}, have been established in variety of applications, like network negotiation, energy trading \cite{ChaBaaKai:2020} and traffic management \cite{LopInnBus:08}, multi-robot systems \cite{ItoHatZhaMat:08}, manufacturing service allocation \cite{KanTanZho:22} and newly in ransomware negotiation \cite{RyaFokHeaAma:22}.
While solving negotiation task, agents must take into account incomplete information and strategically interact with other, human or artificial, agents. Majority of the existing research however assumes negotiation with non-human agents.

Here we consider the simplest bilateral bargaining scenario with incomplete information often found in e-commerce \cite{RenZha:14}. A typical example is two self-interested agents (say, a buyer and a seller) bargaining on some goods or service. As soon as their price preferences differ, agents begin negotiations to achieve a mutually acceptable price. Either agent strives to satisfy own preferences as much as possible, but also has to take into account the opponent’s preferences. Otherwise it is unlikely that an agreement can be reached\footnote{for details on modelling bargaining, see for instance \cite{KenWil:89}.}. Additional aspects of real-life bilateral bargaining to be considered are: i) \textit{multi-attribute negotiation} when agents need to agree on goods/service characterised by several, possibly interrelated, attributes (say price of a product and terms of its delivery); ii) \textit{limited negotiation time} as no agent can deliberate infinitely; iii) \textit{absence of moderator} to coordinate the negotiation, so the agents must reach agreement themselves \cite{Rub:82}.

The negotiation has been widely addressed in diverse fields ranging from economy and sociology to computer science. An amount of works is much too large to survey them here. One can distinguish several main frameworks: game theoretic approach, negotiation protocols approach, evolutionary approach. Existing works however have different limitations preventing them from wide use.
Game theoretic approach \cite{Nas:50}, \cite{Rub:82}, assumes that agents are perfectly rational and have common knowledge.
Negotiation protocols approach, \cite{Rai:82}, needs the clear rules for negotiation, \cite{OsbRub:94}, and the results largely depend on the information available to the agents about each other. Evolutionary approach ,\cite{Ell:97}, being inspired by biological evolution, finds optimal negotiation via trial-error and agents should have access to policy of their opponents and their profits. Some approaches are based on an agent-coordinator responsible for assigning goods or services to agents. This coordinating (or planning) agent uses a negotiation mechanism to find the best share.

We consider a finite horizon bilateral sequential bargaining of two independent self-interested Bayesian decision making (DM) agents facing with incomplete information. The \textbf{key aspects of the targeted solution} are as follows.
\begin{itemize}
	\item \textit{Negotiation.}
	The purpose of negotiation is to enable agents to coordinate their actions/decisions. Thus negotiation is a \textit{means} to achieve coordinated behaviour of the agents. We consider the ability to negotiate an \textit{intrinsic} part of an agent and treat it accordingly.
	The proposed solution allows indirect negotiation via information feedback and further leads to coordinated behaviour without conventional (explicit) negotiation.
	
	\item \textit{Domain-independence.}
	Existing solutions are either of domain-specific, \cite{FicPfe:08}, or domain-independent, \cite{LinKraWilBar:08}.
	The former ones may be more effective, but tailoring them to a new domain may often be useless. The ever-growing number of new applications make domain-tailored solutions less favourable. The considered Bayesian DM agent is inherently domain-independent.
	
	\item \textit{Modelling and learning the opponent.}
	Incomplete knowledge is given by uncertainty regarding the opponent's preferences and behaviour. This uncertainty may prevent agents from reaching mutually beneficial agreement as well as own DM goals. The proposed solution uses Bayesian approach to dynamically learn opponent's model based on observed actions (bids).
	
	\item \textit{Bounded rationality.}
	Assumption on perfect rationality used by game theoretic approach is not valid in real-life tasks. Moreover human agents often behave seemingly irrational due to cognitive or social factors \cite{NeaBaz:92}. Their DM is also influenced by emotional state \cite{WouChaSan:10} and personal traits \cite{Loe:89}: self-interest, altruism,  ability to cooperate.
	The proposed solution is general enough and has already proven to take into account human-like factors \cite{GuyKarLinVil:16}. Thus the approach can serve both an artificial agent and a human.
\end{itemize}

Other important aspect of the negotiation problem concerns \textit{limited deliberation}. Obviously, no agent can bargain indefinitely so the DM policy that is being designed must take that into account. It is hardly possible to set flexible limits on the length of negotiations, but we believe that the established internal feedback complemented by stopping rule can adaptively influence the length of negotiations. A natural decrease of the utility of goods/service over time can also be counteracted by introducing a kind of forgetting \cite{KulKar:84} in the utility function.

\noindent \textbf{Main contributions. }
The paper contributes to research on bilateral bargaining in distributed settings. We propose a \textit{self-interested probabilistic DM agent} maximising expected utility, that is able to purposefully negotiate. The developed agent is domain-independent, can serve to either human or artificial agents and is equipped with the following abilities (which indicate major contributions):
\begin{itemize}
	\item \textit{\textbf{Learning ability.}} To counteract incomplete knowledge and adapt to possible changes of its opponent, the agent is equipped with the learning ability. The algorithm is based on Bayesian approach and learns opponent's model from bargaining history, i.e.~from the bids the opponent proposes during a negotiation. This allows to respect the opponent's dynamics as well as any other related uncertainty, cf.~\cite{BoGaLes:09}.
	
	\item \textit{\textbf{Indirect negotiation.}} A key component of the proposed bargaining agent is a reward function that consists of two components. The first one respects a purely economical individual profit of the agent. The second component expresses degree to which bargaining agents exploit the game potential. It is important to note that the second component i) provides the agent with information feedback; ii) prompts the agents for indirect negotiation, and iii) set limits on the negotiation range. The trade-off between the individual profitability and the game potential is expressed by an agent-specific weight, cf.\cite{HaiGalAnKra:17}. Naturally the opponent equipped with learning ability can model the weight and use this knowledge in next rounds. The weight expresses the agent preferences and partially reflects personal behavioural aspects of human bargaining. The latter opens an avenue for design of automated agents reflecting human traits, \cite{Lin:2012}.
	
	\item \textit{\textbf{Privacy preservation.}} The implicit nature of the resulting distributed interaction does not involve the exchange of any private data or models between players. Therefore the proposed approach fully preserves players' privacy. 	
\end{itemize}
 The proposed solution also allows to incorporate prior knowledge of the opponent though does not require that. The methodology \cite{QuiKarGuy:17} makes it possible to use the available external or domain-specific knowledge to enrich the opponent's model. The paper also compares three types of prior knowledge reflecting typical cases and illustrate its use.

 The paper continues our previous work \cite{HomZugGuy:17} that assumes complete knowledge of the opponent model, which is rarely achievable in real-world applications. Thus the present paper focuses on learning the opponent model as well as on intrinsic motivation to cooperate.
 The last contribution of the work is that we have compared the performance of the proposed bargaining agents to agents employing heuristic models built on the extensive experimental meta-study \cite{Coo:10}.

\noindent \textbf{Related research. } The literature of negotiation constitutes a very large collection, and space limitations prevent it from being presented in its entirety here. Generally there are several models focusing on the explicit negotiation based either on game theory or negotiation theory. The proposed approach considers independent dynamical self-interested DM agents, with learning ability and special reward prompting \textit{\textbf{indirect}} negotiation.
The mentioned features are very practical and up to now missing within the otherwise well-elaborated and important area of the paper. Up to the authors best knowledge there is no similar approach. We use probabilistic models \cite{KarGuy:12} of bargaining agents that interact in a closed-loop and admit Markov decision processes as a modelling methodology, cf.\cite{FagOssLucMil:12}.
The area of agent negotiation and opponent modelling has a lot of achievements, see for instance \cite{FarSieJen:00}, \cite{ZenSyc:98}\cite{WuCheGaoZHe:21}. The comprehensive survey can be found in \cite{BarHenHinJon:16} and in \cite{KirSomKan:20}. The recent paper \cite{BaaKaiGerJonGra:22} discusses main challenges and promises in the area.
Most research on negotiating concerns \emph{static} environments and focused on i) developing utility-based negotiation strategies for rational DM agents, see for instance \cite{FatWooJen:04}, \cite{KroChiMar:21}, and ii) creating agent-moderator helping DM agent in negotiation task, \cite{LinKraWilBar:08}, \cite{LinGevKra:11}, \cite{BoSimMiaShe:08}. So far much less research describe negotiating in dynamic environment, see \cite{FatWooJen:07}, \cite{FenMin:14}. The recent approach \cite{SteKraSar:22} uses a logistic regression for modelling the opponent, that requires collecting significant amount of data for learning and initialisation.
Paper \cite{ZhoChuJinQia:21} uses a similar utility based on the bargaining principles though constructs a subgame that relies on the perfect equilibrium.
 The closely related work, dealt with opponent modelling, is probably \cite{HinTykh:08}. It also employs Bayesian learning but relies on specific structure of preferences and policy of the opponent.
Though work \cite{BoLesSim:10} also focuses on design of negotiation
agents in dynamic and uncertain environments, it relies on a negotiation agent and proposes a set of heuristics to make negotiation decisions.
Our model introduces an intrinsic mechanism that motivates the agent to negotiate while learning opponent's model via Bayesian approach. The resulting bargaining policy is optimal with respect to the resources available and individual preferences of the agents. It can also take into account human factors, which are important whenever human agents are involved.

We illustrate the approach using the Nash Demand Game (NDG)\cite{Nas:50},
 a bilateral bargaining game for two players that should decide how to split given amount of money. The players simultaneously demand a certain portion of the amount they would like to get. The demand of one player is unknown to another one (an opponent). If the players' demands can be satisfied simultaneously, both players get the respective profit. Otherwise, they both get nothing. Despite its seeming simplicity, the NDG is a good model of dynamical resource allocation that achieves coordination without explicit negotiation. It also serves a big challenge for understanding human negotiation.

The remainder of the paper is organized as follows.
Section~\ref{sec:Preliminaries} introduces notations and a mathematical background. Section~\ref{sec:Methodology} formulates the Nash Demand Game as MDP of a single player, introduces heuristic model of the opponent and prior models used in learning. Section~\ref{sec:Experiment} describes and discuss simulated experiments. Section~\ref{sec:discussion} and Section~\ref{sec:Conclusions} summarise the results obtained and outlines future research directions.

\section{Preliminaries} \label{sec:Preliminaries}
This section introduces and recalls necessary notions.
\subsection{General Conventions} \label{sec:GenCon}
    \begin{tabular}{{lp{0.85\linewidth}}}
        $\mathbb{N}$, $\mathbb{R}$      &set of natural numbers, set of real numbers \\
        $x_t\in\mathbf{X}$              &value $x$ from finite set $\mathbf{X}$ at discrete time $t$\\
        $p(x)$            &probability mass function of discrete random variable $x$\\
        $p(x|y)$                        &probability mass function of $x$ conditioned on $y$ \\
        $E[x|y]$                        &the expectation of $x$ conditioned on $y$ \\
    \end{tabular}
Note that no notational distinction is made between a random variable and its realisation.
\subsection{Markov Decision Process} \label{sec:MDP}
We model player's decision making in the NDG via Markov Decision Process (MDP) framework \cite{Put:94}.
MDPs were first introduced and developed in the operations research and economics \cite{Bel:57}. Since that MDP framework has been widely used to describe and solve decision-theoretic problems. MDP allows to capture the underlying stochastics omnipresent in application domain and also allows to respect multiple DM criteria. Typical examples of using MDP framework include medical applications \cite{BogHoePouBouFerMih:06}, predictive maintenance \cite{FenLi:22}, power systems \cite{SonLiuLawDah:00}, more examples see \cite{SchAdaBel:18}.

The overall scenario is as follows. An player interacts with the environment by taking actions to achieve its\footnote{"It" is used as the generic pronoun. A device or an algorithm can be considered as the agent.} DM goal. The player is motivated by a reward it receives after each action taken. A finite state and action MDP is considered.

\begin{definition}[\textbf{MDP}] \label{def:MDP}
  The fully observable MDP is characterised by $\lbrace\S{T},\S{S}, \S{A}, p, R\rbrace$, where $\S{T}=\lbrace 1, 2, ..., N\rbrace, N\in \mathbb{N},$ is a set of {decision epochs}; $\S{S}$ is a finite set of all possible environment states and $\S{A}$ denotes a finite set of all {actions} available to the player. Function $p: \,\S{S}\times\S{S}\times\S{A}\mapsto [0,1]$ is the {transition model} $p(s_{t+1}|s_{t}, a_{t})$ that moves the environment from state $s_{t}\in\S{S}$ to state $s_{t+1}\in\S{S}$ after the agent took action $a_{t}\in \S{A}$; $R: \,\S{S}\times\S{S}\times\S{A}\mapsto \mathbb{R}$ is a real-valued function representing the player's reward $R(s_{t+1},s_{t}, a_{t})$ after taking action $a_{t}\in \S{A}$ in state $s_{t}\in\S{S}$.
\end{definition}

The transition model captures \textit{environment dynamics} and is represented by a family of probability distributions $p(s_{t+1}|s_{t}, a_{t})$, each denotes the probability that at time $t+1$ the environment will move from $s_{t}$ to $s_{t+1}$ when action $a_{t}$ is executed. The state transitions obey Markov property: the distribution over states at time ${t+1}$ is independent of any previous state $s_{t-j}$ and action $a_{t-j}$, $j\leq 1$ for fixed $s_{t}$ and $a_{t}$.

The player's preferences are described by a reward function, $R$. The aim of the player is to choose a sequence of actions in order to maximise the total expected sum of rewards as described in the following section.

\subsection{Optimal Decision Policy} \label{sec:Policy}
The player chooses action $a_{t}\in \S{A}$ based on the \emph{randomised DM rule} $p_{}(a_{t}|s_{t}):\S{S}\mapsto\S{A}$ in each decision epoch $t\in\S{T}$.
A sequence of DM rules forms \emph{DM policy} $\pi_{t,h}$ at time $t$ over decision horizon $h\in \mathbb{N}$, $s_{\tau}\hspace{-3pt}\in\hspace{-2pt}\mathbf{S}, a_{\tau}\hspace{-3pt}\in\hspace{-2pt}\mathbf{A}$:
\begin{equation} \label{eq:DMpolicy}
    \pi_{t,h}\hspace{-3pt} = \hspace{-3pt}\Bigg\{ p_{}(a_{\tau}|s_{\tau})\Big|s_{\tau}, a_{\tau}, \hspace{-3pt} \sum_{\substack{a_{\tau}\in\mathbf{A}}} p_{}(a_{\tau}|s_{\tau})\hspace{-3pt}=\hspace{-3pt}1, \forall s_{\tau}\hspace{-3pt}\in\hspace{-2pt}\mathbf{S}\Bigg\}_{\hspace{-1pt}\tau=t}^{\hspace{-1pt}t+h-1}.
\end{equation}
MDP with finite horizon $h$ evaluates the quality of DM policy by \emph{expected total reward} defined as follows:
\begin{IEEEeqnarray}{lcl}
\label{e:expectation}
    &E&\biggl[\sum\limits_{\tau=t}^{t+h-1} R(s_{\tau+1},s_{\tau}, a_{\tau}) | s_{t} \biggr]=\\
     &&\sum\limits_{\tau=t}^{t+h-1}\hspace{-3pt}\sum\limits_{\substack{s_{\tau+1}\in\mathbf{S} \\s_{\tau}\in\mathbf{S} \\ a_{\tau}\in\mathbf{A}}} {R(s_{\tau+1},s_{\tau}, a_{\tau}) p(s_{\tau+1},s_{\tau}, a_{\tau} | s_{t})},\IEEEnonumber
\end{IEEEeqnarray}
where
\begin{equation}
\label{e:trans_prob}
p(s_{\tau+1}, s_{\tau}, a_{\tau} | s_{t}) = p_{}(s_{\tau+1} | s_{\tau},a_{\tau})p_{}(a_{\tau}|s_{\tau})p(s_{\tau}|s_{t}).\nonumber
\end{equation}

The solution to MDP \cite{Put:94} is a sequence of DM rules, $\Big\{ p_{}^{opt}(a_{\tau}|s_{\tau}) \Big\}_{\tau = t}^{t+h-1}$, that maximises the expected reward (\ref{e:expectation}) and forms the optimal decision policy:
\begin{equation}
\label{eq:optimal_policy}
  \pi_{t, h}^{opt} = \arg\max\limits_{\{\pi_{t, h}\} \in \pmb{\pi}} E \Bigg[ \sum\limits_{\tau=t}^{t+h-1} {R(s_{\tau+1},s_{\tau}, a_{\tau})} | s_{t} \Bigg],
\end{equation}
where $\pmb{\pi}$ is a set of possible DM policies, see \eqref{eq:DMpolicy}.
The optimal policy \eqref{eq:optimal_policy} is computed by dynamic programming algorithm \cite{Bel:57}, \cite{Pow:11}, which requires knowledge of transition model $p_{}(s_{\tau+1} | s_{\tau},a_{\tau})$.

\subsection{Learning Transition Model} \label{sec:LNG}
In bilateral bargaining, the transition model is a model of the opponent, that is, it predicts the opponent's reaction to the player's action. Generally it describes the dynamics of the opponent's decision making.
In real-life tasks, opponent model $p(s_{t+1} | s_{t},a_{t})$ is usually unknown to the player\footnote{it can be partially known or incorrectly specified.}. It reflects the player's knowledge about the behaviour of the opponent. Without lost of generality the model can be assumed time-invariant, i.e. $p(s_{t+1} | s_{t},a_{t})=p(s_{t} | s_{t-1},a_{t-1})$ and can be learned from the observed data.

To simplify the presentation, let us drop out the time index and introduce the following temporary notations: $s'=s_{t+1}$, $s=s_{t}$ and $a=a_{t}$. The transition model then can be written $p(s'|s,a)$\footnote{The new notation is valid within Section \ref{sec:LNG} only.}.

We consider a \textit{parametrised form} of the opponent's model with time-invariant parameter $\theta \in \S{\Theta}$
\begin{equation}
\label{e:parametric}
p(s'|s,a,\theta)= \theta_{s'sa}, \;\;\ \theta_{s'sa} \in \S{\Theta},
\end{equation}
where $\S{\Theta}$ is a set of all possible $\theta$'s and $0\leq\theta_{s'sa}\leq 1$, $\sum_{s'\in\S{S}}\theta_{s'sa}=1$, $\forall (s,a) | s\in\S{S}$ and $a\in\S{A}$.
%
%
%
Thus, parameter $\theta$ in \eqref{e:parametric} is an array defining transition probabilities $\theta_{s'sa}$ that opponent's state in the next time will equal $s'$ whenever the previous state is $s$ and the player takes action $a$. Our aim is to learn parameter $\theta$, \eqref{e:parametric}.

Let the player have belief $b(\theta)$ about the opponent's dynamics expressed via the probability density function of the parameter $\theta$. While interacting with the opponent, the player updates belief about the parameter, $b(\theta)$, to a new value, $b'(\theta)$, given observed transition $(s',s, a)$ as follows, see \cite{Pet:81}:
\begin{equation}\label{e:beliefupdate}
  b'(\theta) \propto b(\theta) p(s'|s, a,\theta)=b(\theta)\theta_{s'sa}.
\end{equation}

Choosing belief $b(\theta)$ in conjugate form of Dirichlet distribution implies that the
posterior (\ref{e:beliefupdate}) induced by Bayes' rule \cite{Pet:81} is
\begin{equation}
\label{e:7}
\mathrm{Dir}(\pmb{\nu},\theta)\propto \prod_{s'sa}\theta_{s'sa}^{\nu_{s'sa}-1}.
\end{equation}
In \eqref{e:7} concentration parameter $\pmb{\nu}>0$ is an array containing occurrences $\nu_{s'sa}>0$ of triples $(s',s,a)$.
Each observation of a triplet $(s',s,a)$ increases the corresponding entry, $\nu_{s'sa}$, by one.

Therefore, after $n \in \mathbb{N}$ observations $\{(s',s,a)\}_{n \in \mathbb{N}}$, update $\nu'_{s'sa}$ contains the actual occurrences of  $(s',s,a)$. Recalling \eqref{e:parametric}, the expectation of (\ref{e:7}) can be interpreted as  Bayesian estimate of unknown parameter $\theta$ based on the observed data (i.e. transitions occurred):
\begin{eqnarray}\label{e:Dirichlet_update}
 E\bigg[ p(s'|s,a,\theta)\bigg| \nu' \bigg]=E\bigg[\theta_{s'sa}| \nu' \bigg]= \frac{\nu'_{s'sa}}{\sum_{s'}{\nu'_{s'sa}}}.
 \end{eqnarray}
Recursive implementation of the prior statistics update is described in \cite{KarBoh:06}.

A real-life dynamic decision making requires an efficient and feasible learning that can be performed online. Markov models belong to the exponential family for which exact estimation is feasible. The estimation and prediction within this family is very simple, especially with the conjugate prior in the form of Dirichlet distribution. The needed update of functions (probability density functions, see~\eqref{e:beliefupdate}) is given by the algebraic recursive update of the finite dimensional sufficient statistics. This clarifies applicability of this learning in combination with decision making.
\section{Methodology}\label{sec:Methodology}

\subsection{MDP Formalisation of Nash Demand Game}\label{sec:ModifiedUG}

The considered repetitive scenario of the game is as follows. Two structurally identical players $\mathcal{A}$ and $\mathcal{B}$ are bargaining on splitting an amount of money $q \in \mathbb{N}$. The roles of both players are the same. In each round, two stages are present: an \emph{action} stage and a \emph{reward} stage. During \emph{action} stage, each player decides how much to claim from the total available amount. The players do not communicate and their interests can be competitive. At \emph{reward} stage, the players announce their demanded shares, observe the demands of their opponents and reward is allocated.
%
Note that in action stage each player has no information about their opponent's demand or preferences. The game runs for a fixed and known number of periods.

Let  $q \in \mathbb{N}$ is a total amount to split. At the beginning of round $t\in \S{T}$, each player $k\in\{\mathcal{A},\mathcal{B}\}$ chooses action $a_{t}^{k}\in\S{A}^{k}$ that is a demanded share of $q$ in the round. The minimum demand equals $1$ and the maximum is $q-1$. If the sum of demands is less than or equal to $q$, both players get what they asked for, otherwise the players get zero reward.

Player's profit in round $t\in \S{T}$ equals the amount of money player receives\footnote[2]{Upper indexes indicate the player whom action or profit belongs to.}:
        \begin{eqnarray}\label{eq:profits}
            \zt{A}&=&a_{t}^{\mathcal{A}}\chi(a_{t}^{\mathcal{A}}, a_{t}^{\mathcal{B}}),\nonumber \\
            \zt{B}&=&a_{t}^{\mathcal{B}}\chi(a_{t}^{\mathcal{A}}, a_{t}^{\mathcal{B}}),
        \end{eqnarray}
where $\zt{A},\zt{B} \in \mathbf{\Z{}}$ are profits of ${\mathcal{A}}$ and ${\mathcal{B}}$ respectively. $\mathbf{\Z{}}=\lbrace 0, 1, 2, ..., q-1 \rbrace$ is a set of possible profits in one game round, and
\begin{equation}\label{eq:chi}
            \chi(a_{t}^{\mathcal{A}},a_{t}^{\mathcal{B}})=
            \begin{cases}
                1 & \mbox{if $a_{t}^{\mathcal{A}}+a_{t}^{\mathcal{B}}\leq q$},\\
                0 & \mbox{if $a_{t}^{\mathcal{A}}+a_{t}^{\mathcal{B}}>q$}.
            \end{cases}
        \end{equation}
The addressed distributed bargaining does not consider communication between the players or any agent-moderator. To find a fully \emph{distributed} solution, the game is described from a point of view of a stand-alone player.
Let us now formulate the discussed bargaining task of a stand-alone player, say player $\mathcal{A}$, as an MDP problem.

\begin{definition} [\textbf{Bargaining as an MDP task}] \label{def:modifiedUG_MDP}
The bargaining scenario is modelled by tuple $\lbrace\S{T},\S{S}, \S{A}, p, R\rbrace$, see Definition \ref{def:MDP}, where
$\S{A}=\lbrace 1, 2, ..., q-1 \rbrace$ is a set of possible actions; $a_{t}^{\mathcal{A}}\in\S{A}$
is \textit{action} of player $\mathcal{A}$, i.e. a portion of $q$ demanded by $\mathcal{A}$ at time $t\in \S{T}$;
$s_{t}=(a_{t-1}^{\mathcal{A}}, a_{t-1}^{\mathcal{B}})\in \S{S}$ is a \textit{state} observed by $\mathcal{A}$ at time $t$ and $p(s_{t+1}|a_{t-1}^{\mathcal{A}}, s_t)$ is a transitional model that describes the state dynamics.
Initial state $s_{1}=(a_{0}^{\mathcal{A}}, a_{0}^{\mathcal{B}})$ is preset to the same demand $a_{0}^{\mathcal{A}}= a_{0}^{\mathcal{B}}=a_{0}$.
\end{definition}
\vspace{5pt}

\noindent\textit{\textbf{Reward as motivation for negotiation.}}
Let reward of player $\mathcal{A}$ be defined as follows:
 \begin{equation}\label{eq:reward}
   R^{\mathcal{A}}_t=a_{t}^{\mathcal{A}}(1-\omega^{\mathcal{A}})\chi(a_{t}^{\mathcal{A}},a_{t}^{\mathcal{B}}) - \omega^{\mathcal{A}}\mid q-(a_{t}^{\mathcal{A}}+a_{t}^{\mathcal{B}})\mid.
\end{equation}
The first term in (\ref{eq:reward}) is a \textit{pure economic profit} of player ${\mathcal{A}}$, cf. \eqref{eq:profits}.
The second term expresses \textit{efficiency of using the game potential} at round $t$, i.e.
whenever $a_{t}^{\mathcal{A}}+a_{t}^{\mathcal{B}}< q$ some amount remains unclaimed and thus lost for the players. The same situation happens when an agreement is not reached and the entire amount $q$ is lost.

Obviously reward \eqref{eq:reward} ensures that, given fixed $a_{t}^{\mathcal{A}}$, player $\mathcal{A}$ will receive the maximum possible reward iff its opponent, ${\mathcal{B}}$, demands $ q-a_{t}^{\mathcal{A}}$.
The proposed form of reward, (\ref{eq:reward}), "connect"  ${\mathcal{A}}$'s action with that of ${\mathcal{B}}$ and thus encourages player ${\mathcal{A}}$ to \emph{indirectly negotiate} with  ${\mathcal{B}}$ during bargaining. The mechanism of dynamic indirect negotiation is as follows.
Each player influences the amount left while their opponent observes this influence and changes their next demand. Let us assume that there is a tendency for some unclaimed amount to remain. Then, if one player has consumed a small portion of it, the other player will observe that and then may increase their demand in the next round. Another situation occurs when the joint claim of the players exceeds the available resources. Then any of the players may step back and reduce their demand in the next round. This behaviour can again lead to a large unclaimed amount and affects the future demands of the players.
In particular, the desire to minimise the unclaimed amount, $\mid q-(a_{t}^{\mathcal{A}}+a_{t}^{\mathcal{B}})\mid$, (\ref{eq:reward}), forces player ${\mathcal{A}}$ to modify the current demand while taking into account the history of the opponent's claims.
By doing so, in each round, each player dynamically \emph{adapts} their demand to the foreseen demands of their opponent, that is indirectly negotiates with the opponent.

\vspace{5pt}
\noindent\textit{\textbf{Weight}} $\omega^{\mathcal{A}}\in [0,1]$ in \eqref{eq:reward} reflects ${\mathcal{A}'s}$ preferences between pure economic gain and exploiting the game's potential. The value $\omega^{\mathcal{A}} = 0$ implies player $\mathcal{A}$ considers pure economic profit only, while in case of $\omega^{\mathcal{A}} = 1$ player $\mathcal{A}$ cares about efficient use of the game potential. The ${\mathcal{A}'s}$ reward (\ref{eq:reward}) thus equals

\begin{equation}\label{eq:reward1}
            R^{\mathcal{A}}_{t}=
            \begin{cases}
                a_{t}^{\mathcal{A}}- \omega^{\mathcal{A}}\left( q-a_{t}^{\mathcal{B}}\right) & \mbox{if $a_{t}^{\mathcal{A}}+a_{t}^{\mathcal{B}}\leq q$},\\
                - \omega^{\mathcal{A}}\left( q-a_{t}^{\mathcal{B}}\right) & \mbox{otherwise}.
            \end{cases}
        \end{equation}
Definition \ref{def:modifiedUG_MDP} and considerations above describe DM of player ${\mathcal{A}}$. Easy to see that the same considerations can be applied to formalise decision making of player ${\mathcal{B}}$.

\vspace{5pt}
The conditional independence of the players' actions given by the game rules and the definition of the state, see Definition \ref{def:modifiedUG_MDP}, imply
\begin{eqnarray}
\label{e:neededM}
p(s_{t+1}|s_{t},a^{\mathcal{A}}_{t})=
p(a^{\mathcal{A}}_{t}|a^{\mathcal{A}}_{t-1},a^{\mathcal{B}}_{t-1})
p(a^{\mathcal{B}}_{t}|a^{\mathcal{A}}_{t-1},a^{\mathcal{B}}_{t-1}).
\end{eqnarray}
From player $\mathcal{A}$ point of view, the first factor in (\ref{e:neededM}) is a part of $\mathcal{A}$'s optimal policy while the second factor models DM of player $\mathcal{B}$ and can be recursively estimated using Bayesian paradigm \cite{Pet:81} as described in Section \ref{sec:LNG}.

\subsection{Heuristic Model of Opponent} \label{sec:models of B}
The proposed approach formalised and solved bilateral dynamic bargaining of learning self-interested player within MDP framework (Section~\ref{sec:Policy}).
To verify the approach we propose a probabilistic bargaining model for non-learning and non-optimising opponent. The model is based on the reported experimental evidence obtained with human-players, see \cite{HenBoy:01}, \cite{Coo:10}.
For simplicity here we consider player ${\mathcal{B}}$ is serving as an opponent to ${\mathcal{A}}$.

Heuristic behaviour of ${\mathcal{B}}$ reflects the dependence of its future demand on the results of the previous round. Once the previous round demands are incompatible, that is $a_{t-1}^{\mathcal{A}}+a_{t-1}^{\mathcal{B}}> q$, player ${\mathcal{B}}$ tends to decrease next demand. If there are unclaimed money left in the previous round, ${\mathcal{B}}$, on the contrary, increases the next demand. The proportion (speed) of demands' increase/decrease may depend on personal traits (i.e. reflect the personality of ${\mathcal{B}}$).

The remainder of this section introduces model that reflects the behaviour of an opposing player, ${\mathcal{B}}$.

\subsubsection{${\mathcal{B}}$ had Low Demand in the Previous Round}\label{sec:low_dem of B}

Consider the previous demand of player ${\mathcal{B}}$ is low, i.e. less than the fair split would have been, $a_{t-1}^{\mathcal{B}}\leq \frac{q}{2}$. The next demand (in sense of its mean value) then depends on the success of the previous round, i.e. whether demands in the previous round were compatible or not. Below we distinguish these two cases and provide the respective probabilistic description of ${\mathcal{B}}$'s actions.
  \begin{enumerate}
     \item [i)]  \textbf{Incompatible Demands ($a_{t-1}^{\mathcal{A}}+a_{t-1}^{\mathcal{B}}> q$)}: ${\mathcal{B}}$ tends to keep its next demand close to the previous one, $a_{t}^{\mathcal{B}}$, as the previous demand of  ${\mathcal{A}}$ was certainly much higher than $a_{t-1}^{\mathcal{B}}$. Thus any further increase could cause players' demands to become incompatible again and implies zero profit. Therefore the new demand of player ${\mathcal{B}}$ can be modelled as follows:
        \begin{equation} \label{eq:model_B_incompart}
          p_{}(a^{\mathcal{B}}_{t}|a_{t-1}^{\mathcal{A}},a_{t-1}^{\mathcal{B}}) \propto \exp\left(-\frac{\left(a^{\mathcal{B}}_{t} - a^{\mathcal{B}}_{t-1}\right)^2}{2\sigma^2}\right)
        \end{equation}
        while $a_{t-1}^{\mathcal{B}}\leq \frac{q}{2}$.\\

     \item [ii)] \textbf{Compatible Demands ($a_{t-1}^{\mathcal{A}}+a_{t-1}^{\mathcal{B}}\leq q$)}: opponent ${\mathcal{B}}$ will proportionally increase the next demand, expecting $\mathcal{A}$ to do the same in order to fully distribute the entire available amount, $q$. In other words player who received less in the previous round would also ask for proportionally less unclaimed money and vice versa. A model of ${\mathcal{B}}$ describing the new demand is then
        \begin{equation}\label{eq:model_B_compart}
          p_{}(a^{\mathcal{B}}_{t}|a_{t-1}^{\mathcal{A}},a_{t-1}^{\mathcal{B}}) \propto \exp\left(-\frac{K}{2\sigma^2}\right),
        \end{equation}
        with $K\hspace{-3pt}=\hspace{-3pt}\left(a^{\mathcal{B}}_{t}-a^{\mathcal{B}}_{t-1}-\frac{a^{\mathcal{B}}_{t-1}}{a^{\mathcal{A}}_{t-1}+
        a^{\mathcal{B}}_{t-1}}(q-a_{t-1}^{\mathcal{A}}-a_{t-1}^{\mathcal{B}})\hspace{-3pt}\right)^2$ while $a_{t-1}^{\mathcal{A}}+a_{t-1}^{\mathcal{B}} \leq q$ and $a_{t-1}^{\mathcal{B}}\leq \frac{q}{2}$.
   \end{enumerate}

\subsubsection{${\mathcal{B}}$ had High Demand in the Previous Round}\label{sec:high_dem of B}
   Now let us consider a situation when the previous demand of ${\mathcal{B}}$ was high, i.e. its value was greater than the fair split would have been, $a_{t-1}^{\mathcal{B}}> \frac{q}{2}$. Then ${\mathcal{B}}$ decreases/increases demand while keeping own share proportional to the previous round in order to fully distribute the entire amount. A player who received less in the last round would ask for less of proportionally less unclaimed money and vice versa. Then a model of ${\mathcal{B}'s}$ new demand has the same form as (\ref{eq:model_B_compart}).

\subsection{Prior Models Used in Learning}\label{model_B_opt}
Our approach considers decision making of the player in question, ${\mathcal{A}}$, who models behaviour of the opponent, ${\mathcal{B}}$, and optimises own demand in order to maximise the accumulated profit. The ability to accurately predict the opponent's behaviour significantly affects the success of ${\mathcal{A}}'s$ decision making, (\ref{e:neededM}).
To learn a model of the opponent, ${\mathcal{A}}$ follows the approach described in Section \ref{sec:LNG}.
It exploits knowledge available in the form of a parameter prior that quantifies ${\mathcal{A}'s}$  belief about dynamics of the opponent, ${\mathcal{B}}$. Following Bayesian paradigm this prior will be gradually updated with new data accumulated, see Section \ref{sec:LNG} and \cite{KarBoh:06}. The choice of prior model is important, especially when a number of game rounds is limited. In implementation we use three prior models reflecting different knowledge ${\mathcal{A}}$ about ${\mathcal{B}}$:
\begin{itemize}
	\item \textit{a uniform prior distribution}. This model is used when ${\mathcal{A}}$ has little or no knowledge about the dynamics of ${\mathcal{B}}$
	
	\item \textit{prior model describing "rational" heuristic}, see Section \ref{sec:models of B}. It is used when non-optimising ${\mathcal{B}}$ follows some heuristic and does not optimise. In that case prior model has the same structure as (\ref{eq:model_B_incompart}) or (\ref{eq:model_B_compart}), but with different (larger) standard deviation $\sigma$.

	\item \textit{pre-trained prior model.} The third way of building an a priori model mimics the natural learning process of human players, where the player first gathers some knowledge about the opponent's playing style and then updates this knowledge during the game.
	Practically it means  we run game for $30$ preliminary rounds and player ${\mathcal{A}}$ built prior model of ${\mathcal{B}}$ based on the data obtained during these rounds. This way of building prior is used whenever the both players optimise and learn.
\end{itemize}

\section{Simulated Experiments} \label{sec:Experiment}
The proposed approach is illustrated with the Nash demand game, described in Section \ref{sec:ModifiedUG}, using simulated examples\footnote{The examples were implemented in MATLAB, The MathWorks, Inc.}.

We selected the most representative experiments from a much wider set of the experiments differing in the number of rounds and horizons. The selected experiments are long enough to perform learning (because very short runs will not be sufficient to learn the models used), while longer runs will add no significant information about the results.
\paragraph*{\textbf{Goal of the experiments}}
The goal was to analyse the impact of the proposed distributed solution and indirect negotiation and to verify that player employing the proposed DM policy is capable of achieving better results than heuristic player playing the same role.
The main objectives of the performed experiments are:
\begin{itemize}
  \item illustrate the distributed DM approach in repetitive bargaining;
  \item show that the proposed form of the reward function leads to an indirect negotiation and to a coordinated course of actions of both players, that is, to a more efficient allocation of the available limited resources;
  \item demonstrate influence of weight $\omega$ in \eqref{eq:reward}
  \item show that DM policy with indirect negotiation brings higher profit to every player compare to the heuristic model.
\end{itemize}

\paragraph*{\textbf{Common settings of the experiments}}

Each game has $60$ rounds and optimisation horizon $h \in \mathbb{N}$ equals $10$ game rounds. The amount of money that players can split (if they reach an agreement) is $q=10$ CZK per round. The reward (\ref{eq:reward}) is evaluated for the optimal policy (\ref{eq:optimal_policy}) resulted from the dynamic programming \cite{Bel:57}. The initial state of each player $s_{1}=(a_{0}^{\mathcal{A}}, a_{0}^{\mathcal{B}})$ is preset to $a_{0}^{\mathcal{A}}= a_{0}^{\mathcal{B}}=3$.

The simulation is performed for $11$ different values of weight $\omega$, (\ref{eq:reward}). Weight ($0\leq\omega\leq 1$) expresses a trade-off between the individual profitability and efficiency of using the game resources. It thus reflects the extent to which the player is negotiating. Zero value of $\omega$ in (\ref{eq:reward}) models the situation when the player is interested only in economic profit. Other values of $\omega$ ($0<\omega\leq 1$) correspond to cases when the player maximises the personal profit while minimising the unclaimed amount of money.

\paragraph*{\textbf{Experiments performed}}
The players used in the simulation are artificial agents with either heuristic DM model (see Section~\ref{sec:models of B}) or proposed DM policy that optimises reward (\ref{eq:reward}), see Section \ref{sec:Policy}.
In each game at least one of the players uses the observed behaviour to update the opponent's model, see Section~\ref{sec:LNG}. In order to display behaviour of our bargaining model, five typical cases were considered:

\begin{description}
\item[\hspace{-2ex}\textbf{Test~1 }:] Both players are non-learning. The player in question, $\mathcal{A}$, is of the MDP type and uses the proposed DM policy  optimising (\ref{eq:reward}). Its opponent, $\mathcal{B}$,  behaves heuristically, see Section~\ref{sec:models of B}.

  \item[\hspace{-2ex}\textbf{Test~2 }: ] This case is similar to Test~1 but player $\mathcal{A}$ dynamically learns the opponent's model.

  \item[\hspace{-2ex}\textbf{Test~3 }: ] Both players are of the MDP type and non-learning. They have  no knowledge of their opponent and do not model it either (i.e. they use uniform model).

  \item[\hspace{-2ex}\textbf{Test~4 }: ] Both players are of the MDP type, and use  having non-informative prior for learning, see Section~\ref{sec:LNG}.

  \item[\hspace{-2ex}\textbf{Test~5 }: ] This case is similar to Test~4, but the players use informative priors (i.e. opponent model trained during the preliminary phase).
\end{description}

\paragraph*{\textbf{Approach verification}} The players have played the game repeatedly with different settings.
The results are summarised in graphs depicting \emph{individual cumulative profits} of the players, total profit of the game, and success rates of game depending on the value of parameter $\omega$. The \emph{success rate} is defined as a number of game rounds in which the players' demands were compatible and thus satisfied. In other words, the value of the success rate shows how successfully the players collaborated, i.e. respected the opponent's actions. High values indicate high collaboration.
 The results show minimum, mean and maximum values of the individual cumulative profits and the game success rate. Note that
\begin{itemize}
  \item The maximum success rate does not necessarily imply the maximum total profit of the game.
  \item Compatibility of the players' claims does not guarantee zero unclaimed amount in the game.
  \item It is not guaranteed either that the maximum profit will be obtained for the same value of the weight $\omega$. Thus the total maximum (minimum) profit of the game is \textit{not} equal to the sum of the individual maximum (minimum) profit of the players.
\end{itemize}

\subsection{Test~1: $\mathcal{A}$ is a non-learning MDP player, $\mathcal{B}$ behaves heuristically} \label{sec:Exp1}
Player $\mathcal{B}$, behaves according to the heuristic model \eqref{eq:model_B_incompart}, \eqref{eq:model_B_compart} with $\sigma^2=1$.

Player $\mathcal{A}$ is of MDP type and uses DM policy \eqref{eq:optimal_policy} that optimises reward \eqref{eq:reward}. In optimisation $\mathcal{A}$ uses model $p(s_{t+1}|s_{t}, a_t)$ having structure of the heuristic model, see Section~\ref{sec:models of B}, but with different parameter $\sigma=3$. This imitates a situation when $\mathcal{A}$ has partial or vague knowledge of the opponent.

\begin{figure}[!ht]
  \begin{minipage}{0.49\textwidth}
    \includegraphics[trim=0.5cm 7cm 14cm 8cm,width=0.31\linewidth]{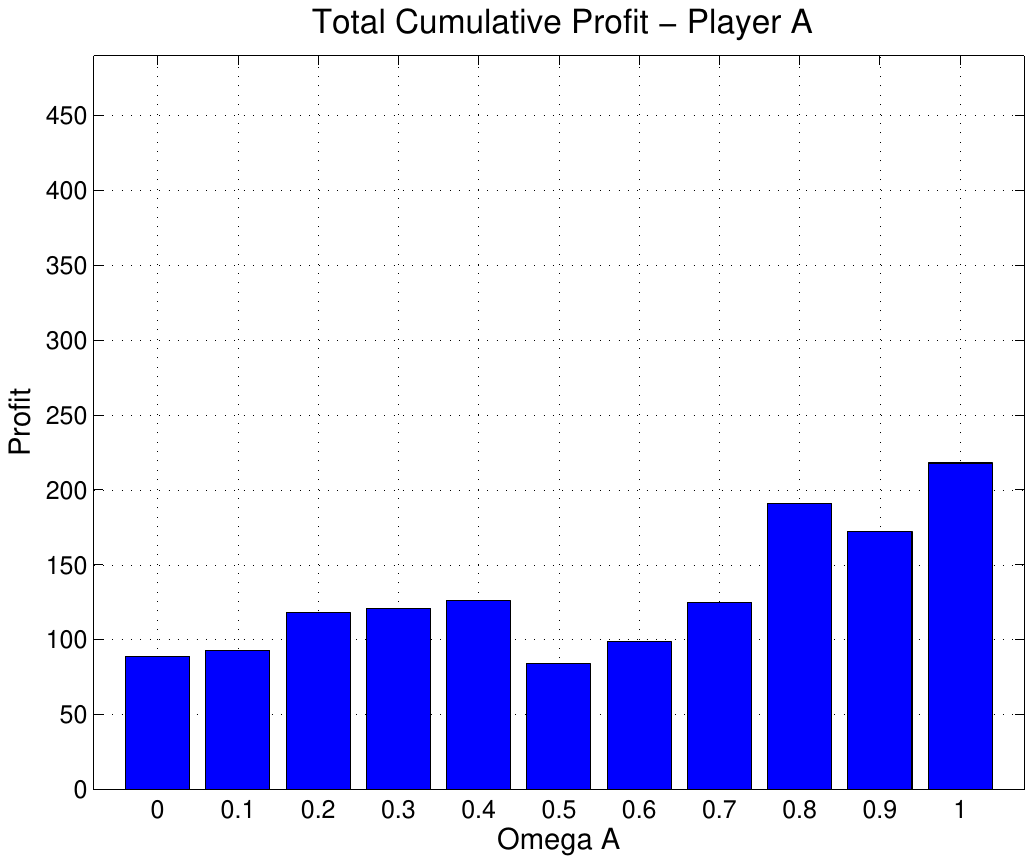}
      \caption{Test~1 - $\mathcal{A}$'s cumulative profit on weight $\omega^{\mathcal{A}}$}\label{fig:exp1_Player A}
 \end{minipage}
 \hfill
  \begin{minipage}{0.49\textwidth}
    \includegraphics[trim=0.5cm 7.5cm 14cm 7cm,width=0.31\linewidth]{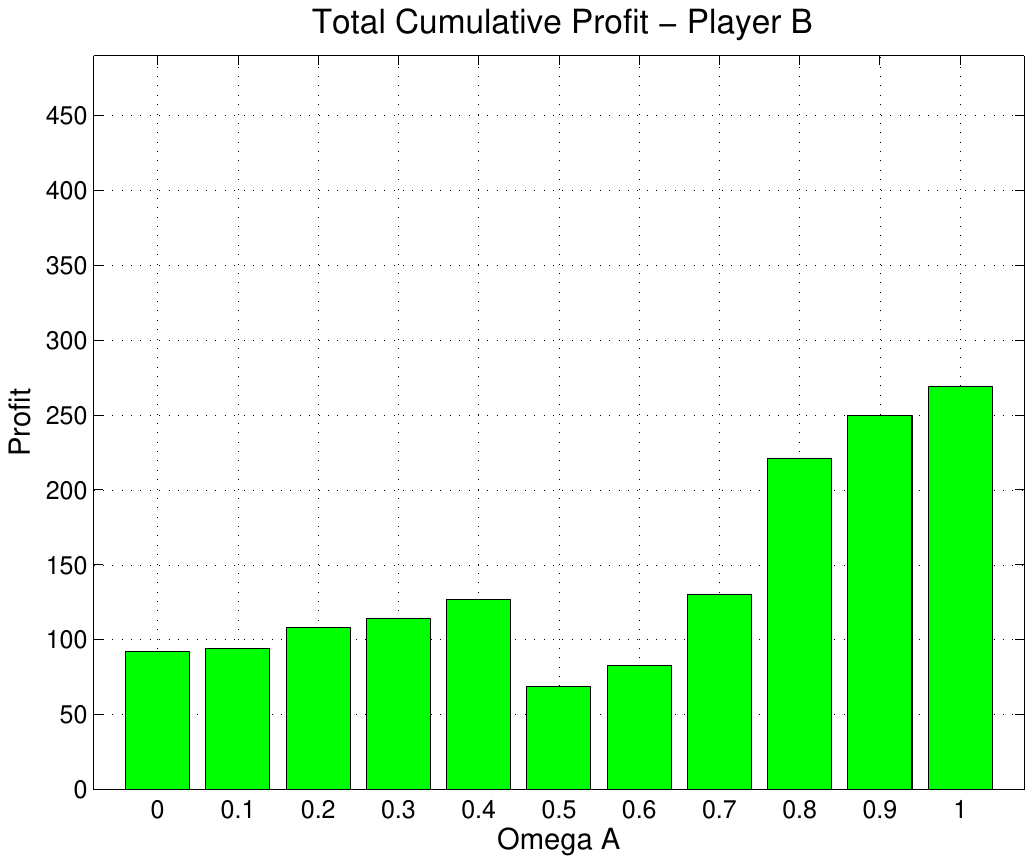}
    \caption{Test~1 - $\mathcal{B}$'s cumulative profit on weight $\omega^{\mathcal{A}}$}\label{fig:exp1_Player B}
 \end{minipage}
\end{figure}

\begin{figure}[!ht]
  \begin{minipage}{0.49\textwidth}
    \includegraphics[trim=0.5cm 7.5cm 14cm 7cm,width=0.31\linewidth]{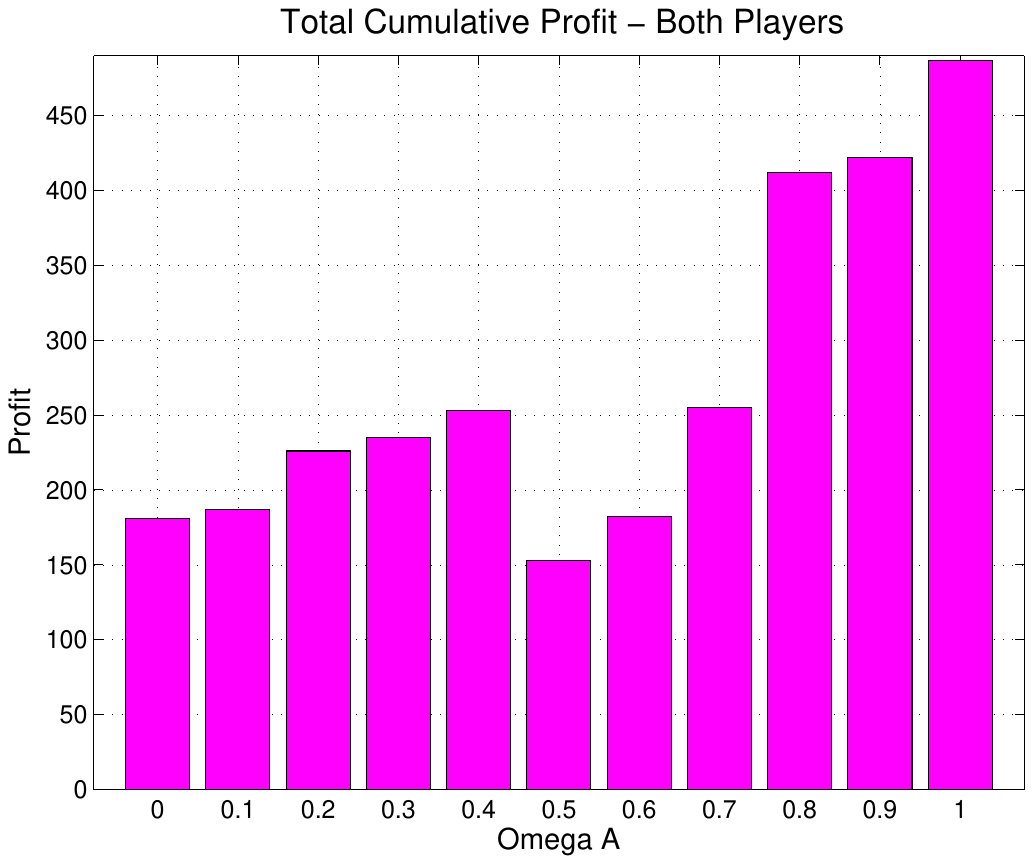}
      \caption{Test~1 - Total Profit of Players.}\label{fig:exp1_Total}
 \end{minipage}
 \hfill
  \begin{minipage}{0.49\textwidth}
    \includegraphics[trim=0.5cm 7cm 14cm 7cm,width=0.31\linewidth]{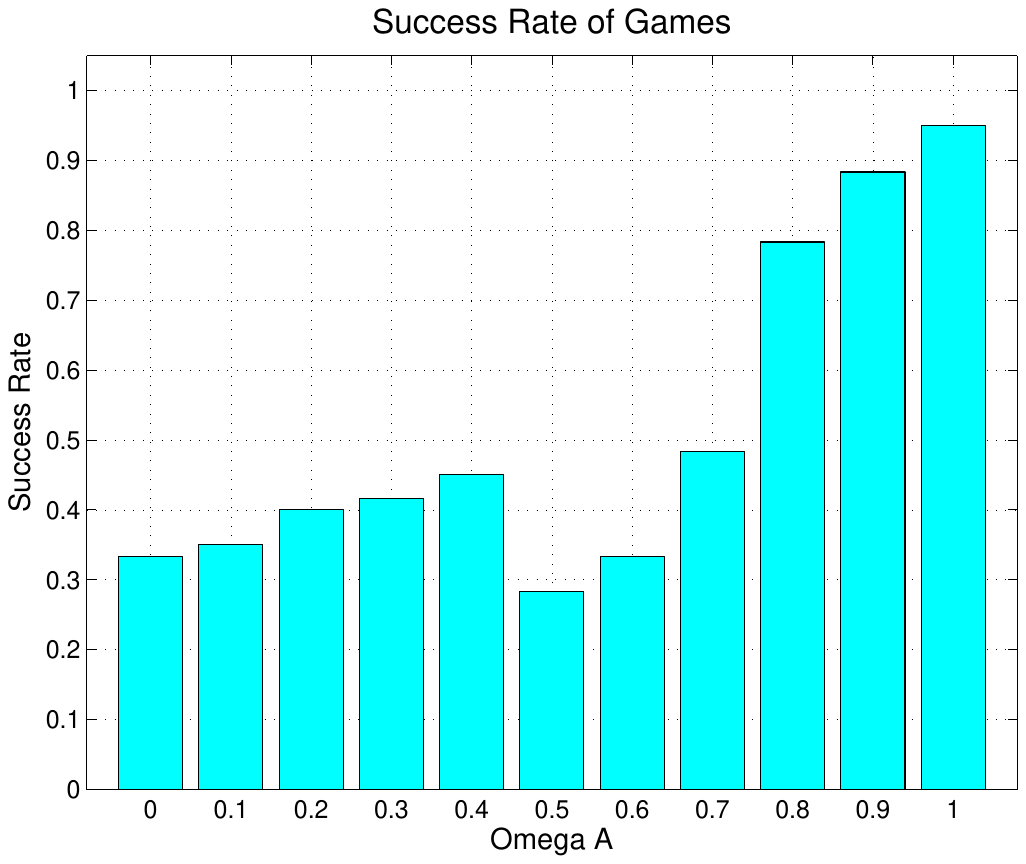}
    \caption{Test~1 - Success Rate of the Games.}\label{fig:exp1_Success}
 \end{minipage}
\end{figure}

Cumulative profits of the players $\mathcal{A}$ and $\mathcal{B}$ are shown in Figure~\ref{fig:exp1_Player A} and Figure~\ref{fig:exp1_Player B}. Total cumulative profit and success rate of the game as a function of parameter $\omega^{\mathcal{A}}$ are shown in  Figure~\ref{fig:exp1_Total} and Figure~\ref{fig:exp1_Success}.

The players are successful in more than 51\% of the rounds on average.
The results show influence of parameter $\omega^{\mathcal{A}}$ on profit: the higher the parameter, the higher the profits of individual players and the higher the total profit of the game. This indicates a positive effect of the second term \eqref{eq:reward}, which prompts $\mathcal{A}$ to \textit{indirectly negotiate} with $\mathcal{B}$ by minimising the unclaimed amount in each round. As a result the players start to implicitly cooperate.

 The results show the saddle value of parameter $\omega^{\mathcal{A}} = 0.5$ that provides the minimum values of $\mathcal{A}$'s profit and success rate of the game. The maximum is reached for $\omega^{\mathcal{A}} = 1$. Obviously, optimising player $\mathcal{A}$ earned slightly less on average than non-optimising player $\mathcal{B}$. It could be because player $\mathcal{B}$ used fixed decision making rules and $\mathcal{A}$ had to adapt to that.

\begin{table}
\caption{Test~1: Player $\mathcal{A}$ optimises but not learn. Player $\mathcal{B}$ follows the heuristic model \eqref{eq:model_B_incompart},\eqref{eq:model_B_compart}.}\label{tab:exp1_table}
\begin{center}
\begin{tabular}{|l|c|c|c|c|}
\hline
 \multicolumn{4}{|c|}{Cumulative profit} & Success rate  \\
 \hline
 & $\mathcal{A}$ & $\mathcal{B}$ & Total & \%\\
\hline
 Min. value & 84.00  & 69.00  & 153.00  & 28.00  \\
 Mean value &130.55 & 141.55  & 272.09  & 51.52  \\
 Max. value & 218.00  & 269.00  & 487.00  & 95.00  \\
\hline
\end{tabular}
\end{center}
\end{table}

\subsection{Test~2: $\mathcal{A}$ optimises and learns, $\mathcal{B}$ behaves heuristically}\label{Exp2}

This experiment is similar to Test~1, see Section~\ref{sec:Exp1}, i.e. player $\mathcal{B}$ behaves accordingly to the heuristic model, Section~\ref{sec:models of B}, and player $\mathcal{A}$ uses optimal DM policy minimising the proposed reward, \eqref{eq:reward}. Unlike Test~1, player $\mathcal{A}$ is learning. $\mathcal{A}$ considers a uniform prior as ${\mathcal{B}}$'s transition model and dynamically updates it via data gathered, see Section~\ref{model_B_opt}.

\begin{figure}[!ht]
  \begin{minipage}{0.49\textwidth}
    \includegraphics[trim=0.5cm 7.5cm 14cm 7cm,width=0.31\linewidth]{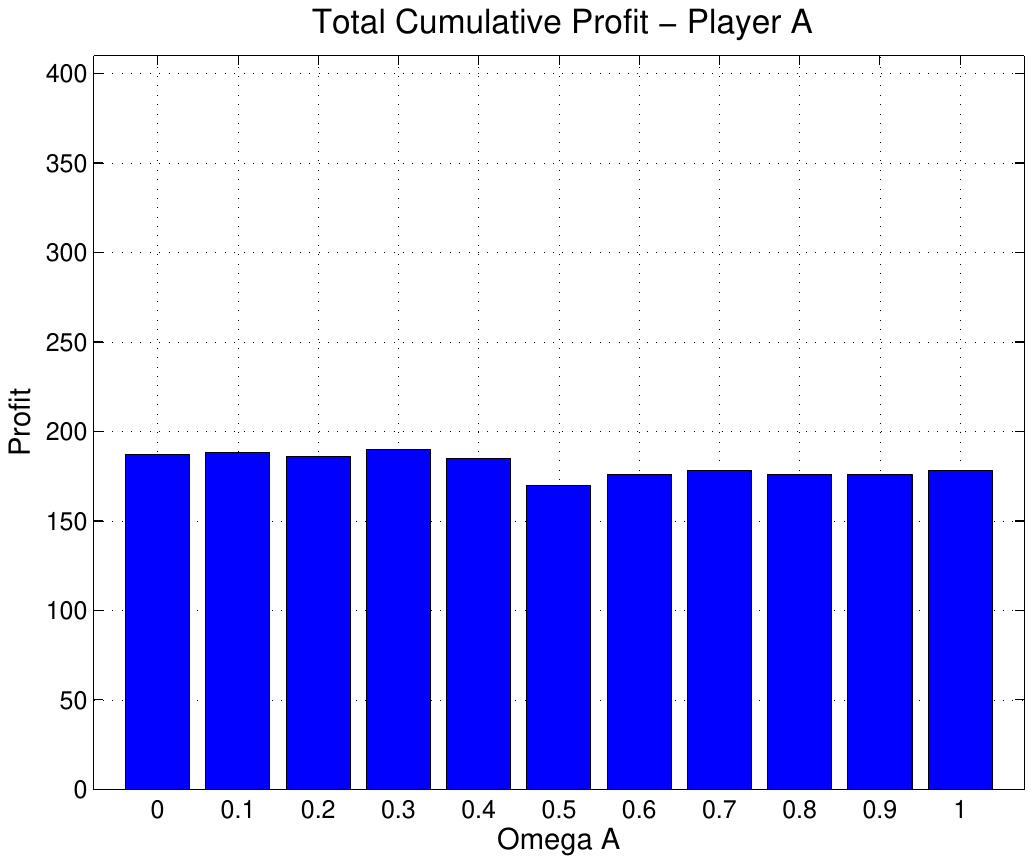}
      \caption{Test~2 - $\mathcal{A}$'s cumulative profit on weight  $\omega^{\mathcal{A}}$.}\label{fig:exp2_Player A}
 \end{minipage}
 \hfill
  \begin{minipage}{0.49\textwidth}
    \includegraphics[trim=0.5cm 7cm 14cm 7cm,width=0.31\linewidth]{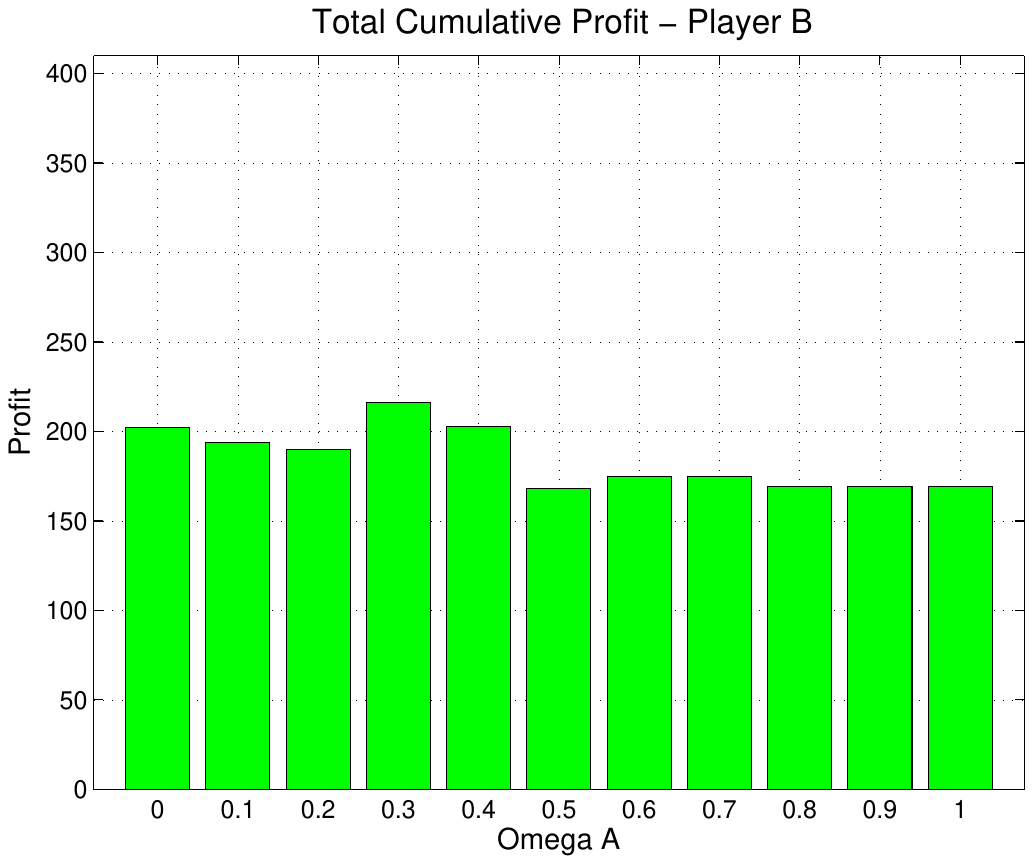}
    \caption{Test~2 - $\mathcal{B}$'s cumulative profit on weight  $\omega^{\mathcal{A}}$.}\label{fig:exp2_Player B}
 \end{minipage}
\end{figure}

\begin{figure}[!ht]
  \begin{minipage}{0.49\textwidth}
    \includegraphics[trim=0.5cm 7cm 14cm 7cm,width=0.31\linewidth]{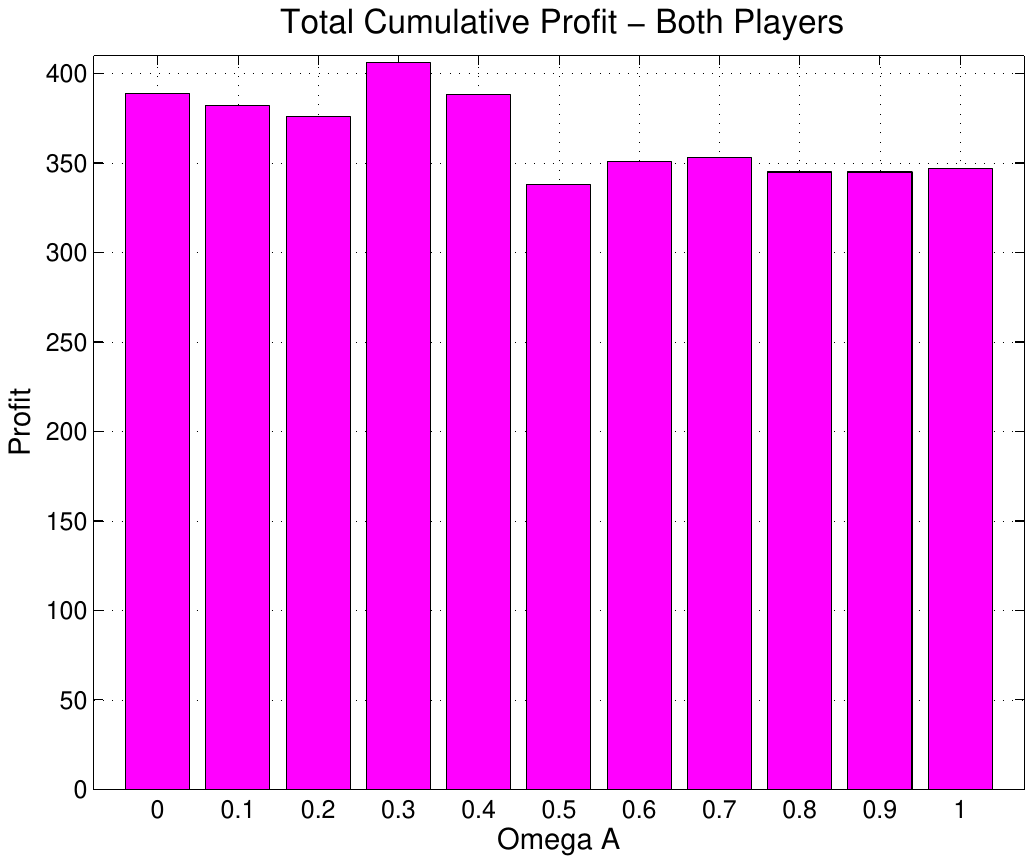}
      \caption{Test~2 - Overall profit of the players.}\label{fig:exp2_Total}
 \end{minipage}
 \hfill
  \begin{minipage}{0.49\textwidth}
    \includegraphics[trim=0.5cm 7cm 14cm 7cm,width=0.31\linewidth]{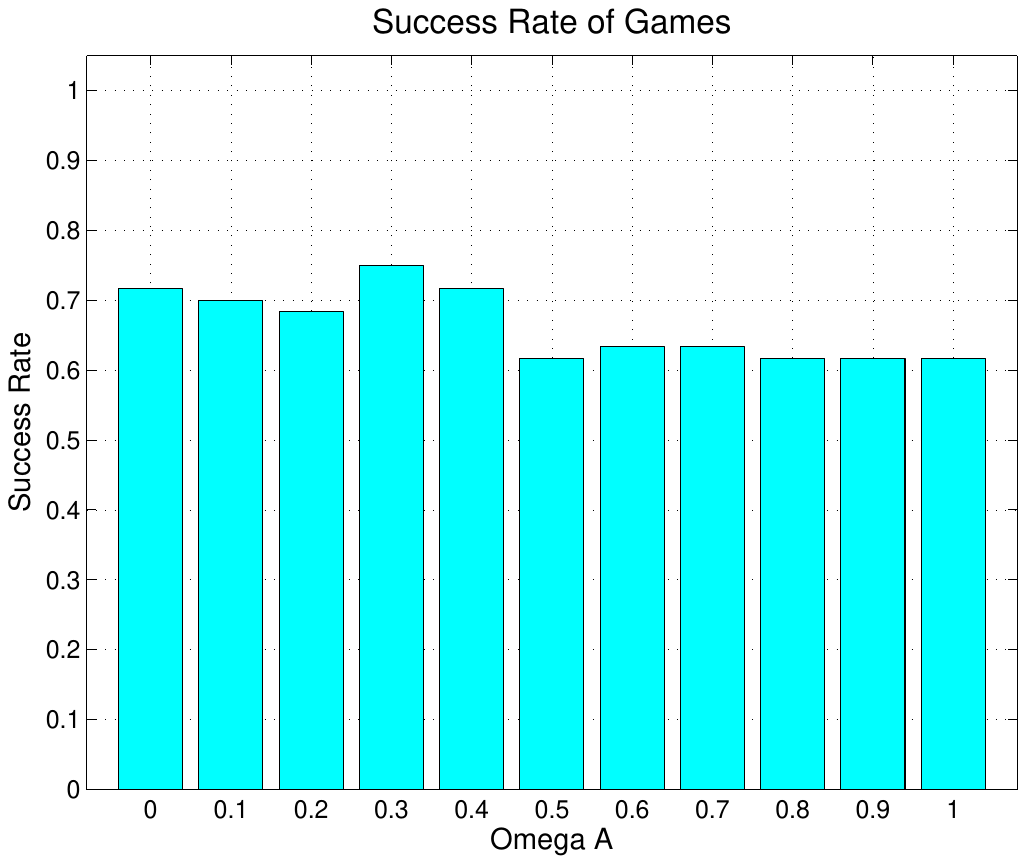}
    \caption{Test~2 - Success rate of the games.}\label{fig:exp2_Success}
 \end{minipage}
\end{figure}

Cumulative profit and success rate obtained in Test~2 are shown in  Figures~\ref{fig:exp2_Player A}-\ref{fig:exp2_Success} and Table~\ref{tab:exp2_table}. Obviously the learning has a positive impact on the game results. On average, the players are successful in more then 66\% of all rounds - the average success rate is about 15\% higher than in Test~1, as is the cumulative profit. The minimum values for individual profits and overall success rate are significantly higher cf. Table~\ref{tab:exp1_table}. On the other hand, their maximum values have noticeably decreased. The players have similar individual profits and their values weakly depend on parameter $\omega^{\mathcal{A}}$.

\begin{table}
\caption{Test~2: Player $\mathcal{A}$ optimises and learns, player $\mathcal{B}$ follows the heuristic model \eqref{eq:model_B_incompart}, \eqref{eq:model_B_compart}.}\label{tab:exp2_table}
\begin{center}
\begin{tabular}{|l|c|c|c|c|}
\hline
 \multicolumn{4}{|c|}{Cumulative profit} & Success rate  \\
 \hline
 & $\mathcal{A}$ & $\mathcal{B}$ & Total & \%\\
\hline
 Min. value  & 170.00  & 168.00  & 338.00  & 62.00  \\
 Mean value  & 180.91  & 184.55  & 365.45  & 66.36  \\
 Max. value  & 190.00  & 216.00  & 406.00  & 75.00 \\
\hline
\end{tabular}
\end{center}

\end{table}

\subsection{Test~3: Both players optimise but none learns}\label{Exp3}
This experiment considers both players are of MDP type and select DM policy maximising reward \eqref{eq:reward}. However neither of the players is learning. They use a fixed uniform model (see Section~\ref{model_B_opt}) that models the situation when there is no information about the opponent.

\begin{figure}[!ht]
  \begin{minipage}{0.49\textwidth}
    \includegraphics[trim=1.5cm 7cm 13.5cm 7cm,width=0.32\linewidth]{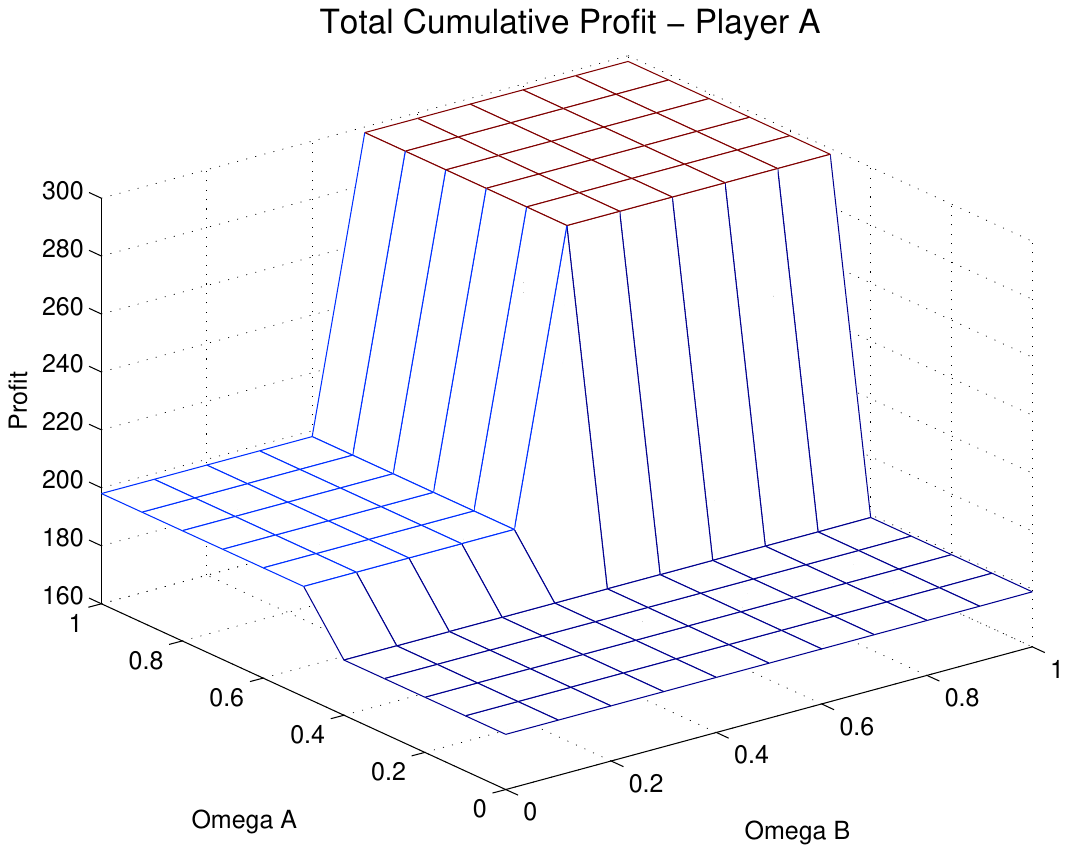}
      \caption{Test~3 - $\mathcal{A}$'s cumulative profit in dependence on weights  $\omega^{\mathcal{A}}$ and $\omega^{\mathcal{B}}$.}\label{fig:exp3_Player A}
 \end{minipage}
 \hfill
  \begin{minipage}{0.49\textwidth}
    \includegraphics[trim=1.5cm 7cm 13.5cm 7cm,width=0.32\linewidth]{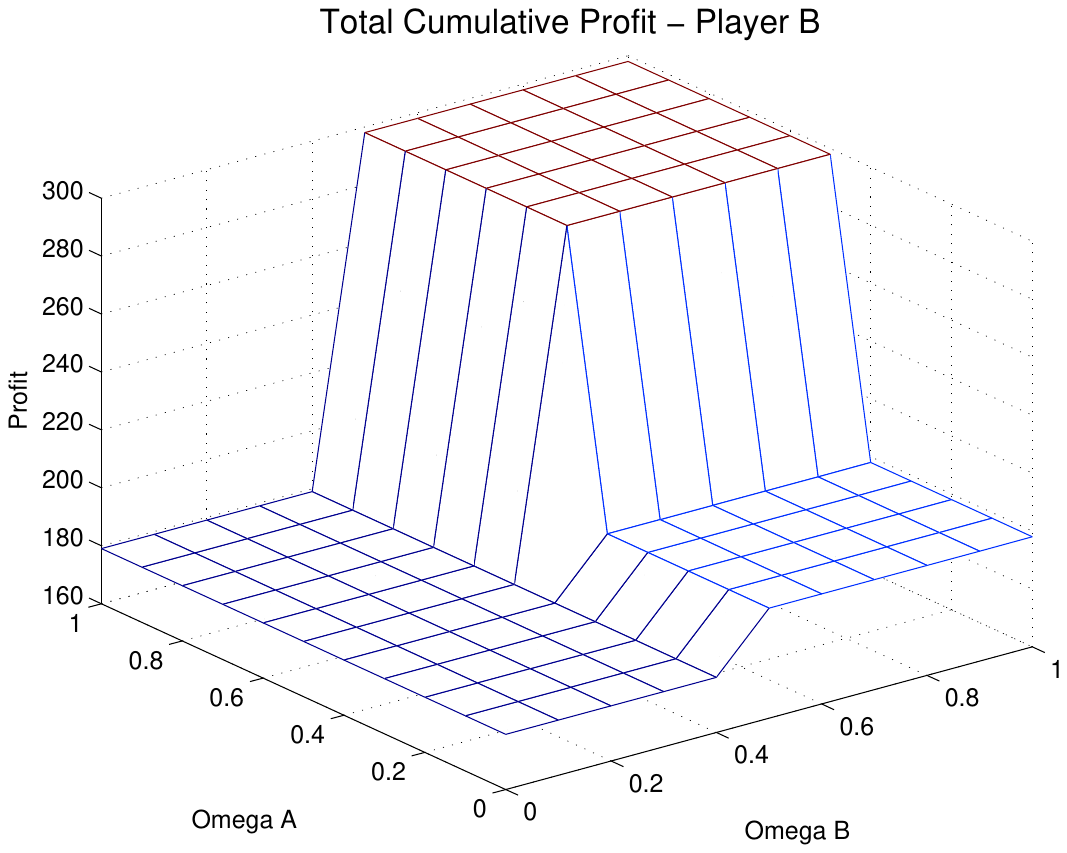}
    \caption{Test~3 -$\mathcal{B}$'s cumulative profit in dependence on weights  $\omega^{\mathcal{A}}$ and $\omega^{\mathcal{B}}$}\label{fig:exp3_Player B}
 \end{minipage}
\end{figure}

\begin{figure}[!ht]
  \begin{minipage}{0.49\textwidth}
    \includegraphics[trim=1.5cm 7cm 13.5cm 7cm,width=0.32\linewidth]{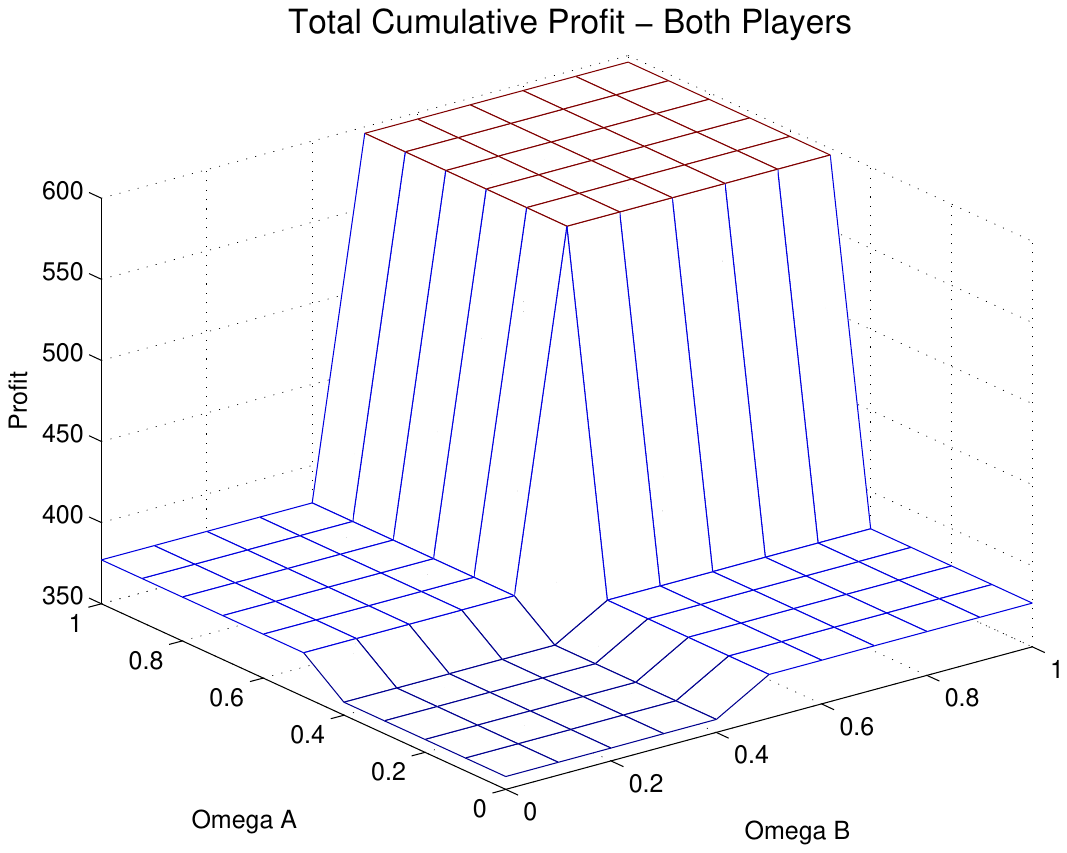}
      \caption{Test~3 - Overall profit of the players.}\label{fig:exp3_Total}
 \end{minipage}
 \hfill
  \begin{minipage}{0.49\textwidth}
    \includegraphics[trim=1.5cm 7cm 13.5cm 7cm,width=0.32\linewidth]{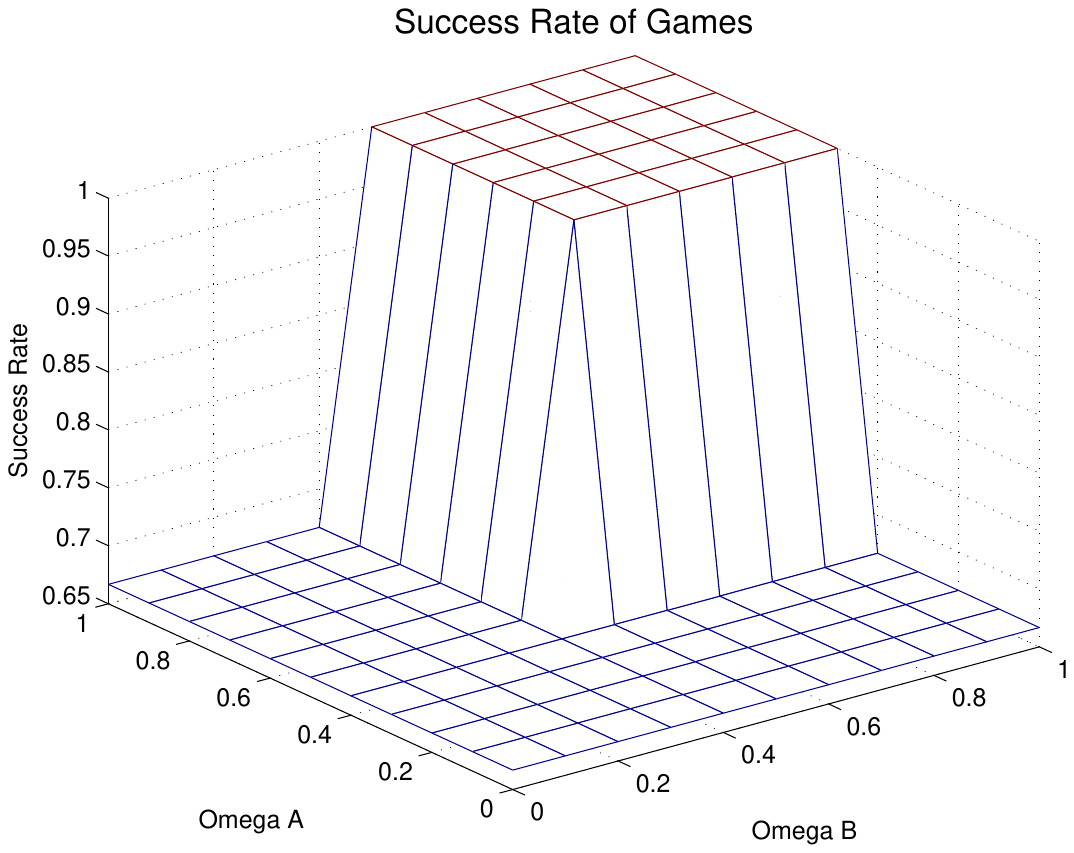}
    \caption{Test~3 - Success rate of the games.}\label{fig:exp3_Success}
 \end{minipage}
\end{figure}

Cumulative profits and success rate of the game vs. parameters $\omega^{\mathcal{A}}$ and $\omega^{\mathcal{B}}$ are shown in Figures~\ref{fig:exp3_Player A}-\ref{fig:exp3_Success} and Table~\ref{tab:exp3_table}.

The results illustrate positive impact of i) optimal bargaining compare to heuristic behaviour, cf. results of Test~1 and Test~2 and ii) proposed reward \eqref{eq:reward} that prompts on indirect negotiation. Even with non-informative prior knowledge, the players get higher profit. If the players' weights are $\omega^{\mathcal{A}}\geq 0.5$ and $\omega^{\mathcal{B}}\geq 0.5$ the success rate is $100\%$ and overall game profit gained is close to the maximum possible ($600$ CZK), see Table \ref{tab:exp3_table}. By other words: when the players care about the optimal allocation of the resources (by assigning high weights to the second term in reward \eqref{eq:reward}), the bargaining is more profitable. On average, the players are successful in more than $76\%$ of all rounds.

\begin{table}
\caption{Test~3: Both players optimise but none learns.}\label{tab:exp3_table}
\begin{center}
\begin{tabular}{|l|c|c|c|c|}
\hline
 \multicolumn{4}{|c|}{Cumulative profit} & Success rate  \\
 \hline
 & $\mathcal{A}$ & $\mathcal{B}$ & Total & \%\\
\hline
 Min. value  & 179.00  & 179.00  & 504.00  & 67.00  \\
 Mean value  & 219.12  & 219.12  & 438.23  & 76.58  \\
 Max. value  & 298.00  & 298.00  & 596.00  & 100.00  \\
\hline
\end{tabular}
\end{center}
\end{table}

\subsection{Test~4: Players optimise and learn with uniform prior}\label{Exp4}

This experiment is similar to Test~3, i.e. both players are of MDP type and maximise reward \eqref{eq:reward}. Unlike Test~3, the ability to learn the opponent's model has been added to the players. The agents dynamically enhance their non-informative (uniform) priors based on the data observed during the game. Thus each player i) learns their opponent; ii) searches for optimal demand; iii) indirectly negotiates via minimising unshared resources.

\begin{figure}[!ht]
  \begin{minipage}{0.49\textwidth}
    \includegraphics[trim=1.5cm 7cm 13.5cm 7cm,width=0.32\linewidth]{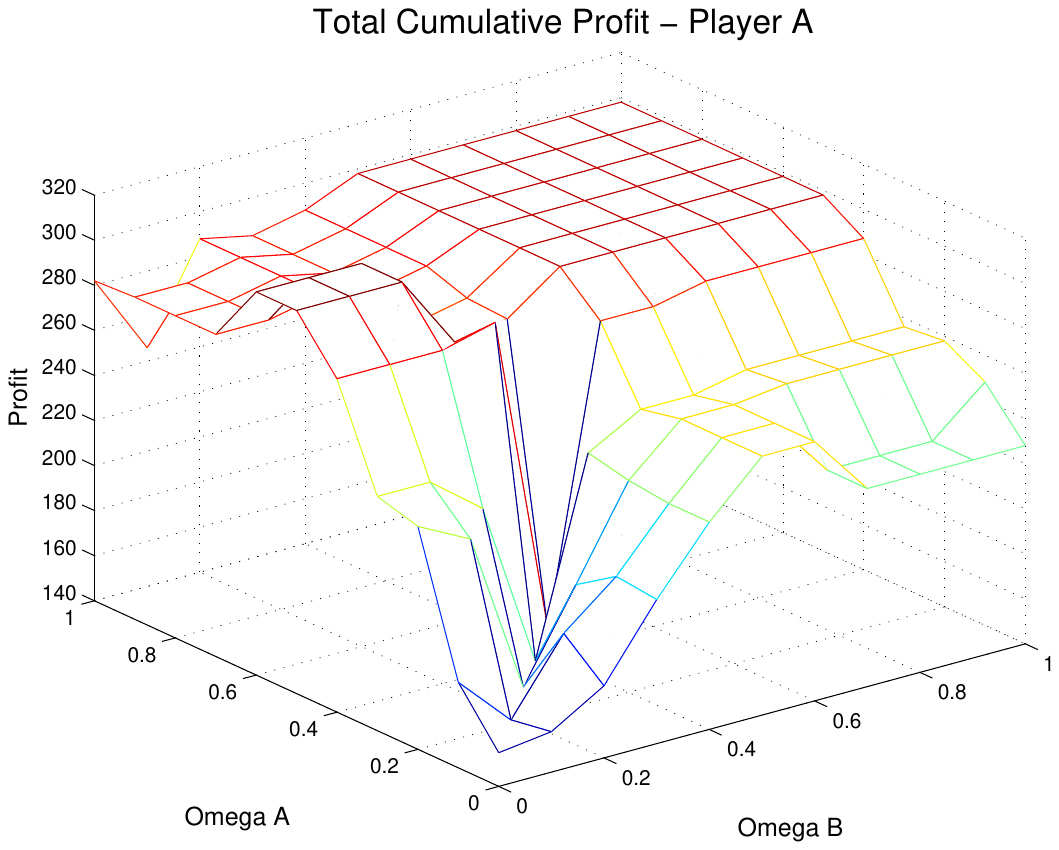}
      \caption{Test~4 - $\mathcal{A}$'s cumulative profit in dependence on weights  $\omega^{\mathcal{A}}$ and $\omega^{\mathcal{B}}$.}\label{fig:exp4_Player A}
 \end{minipage}
 \hfill
  \begin{minipage}{0.49\textwidth}
    \includegraphics[trim=1.5cm 7cm 13.5cm 7cm,width=0.32\linewidth]{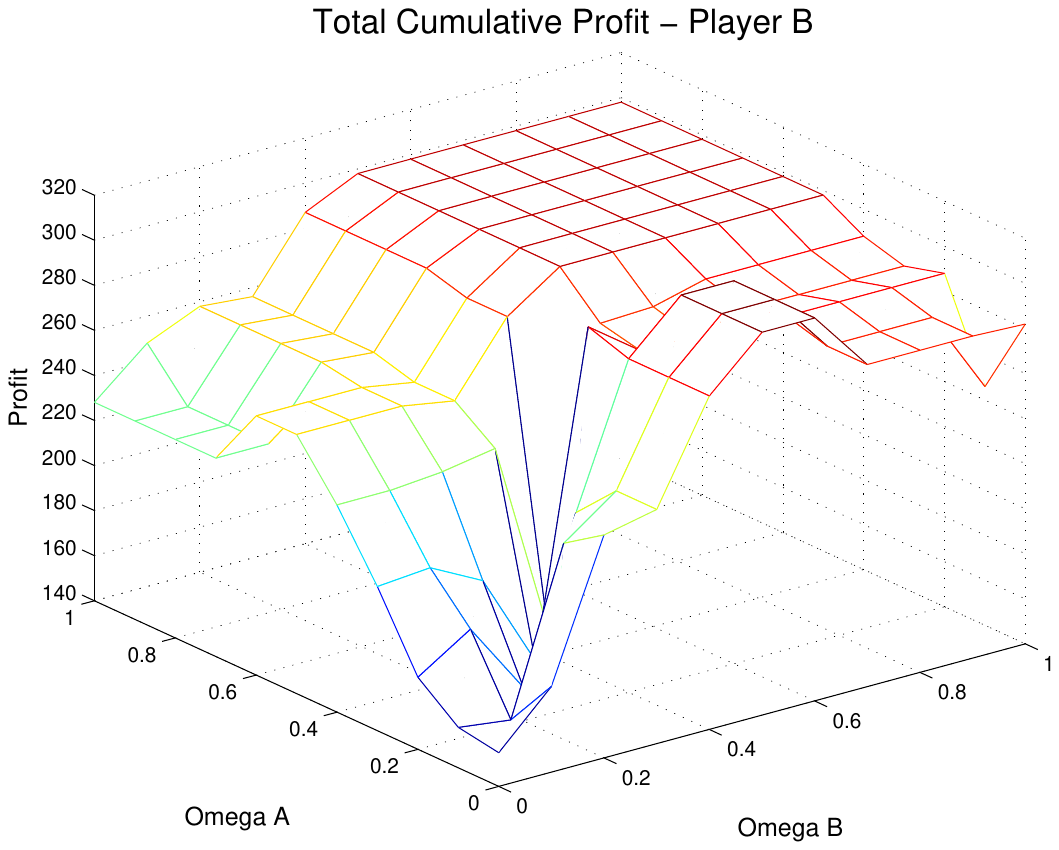}
    \caption{Test~4 - $\mathcal{B}$'s cumulative profit in dependence on weights  $\omega^{\mathcal{A}}$ and $\omega^{\mathcal{B}}$.}\label{fig:exp4_Player B}
 \end{minipage}
\end{figure}

\begin{figure}[!ht]
  \begin{minipage}{0.49\textwidth}
    \includegraphics[trim=1.5cm 7cm 13.5cm 7cm,width=0.32\linewidth]{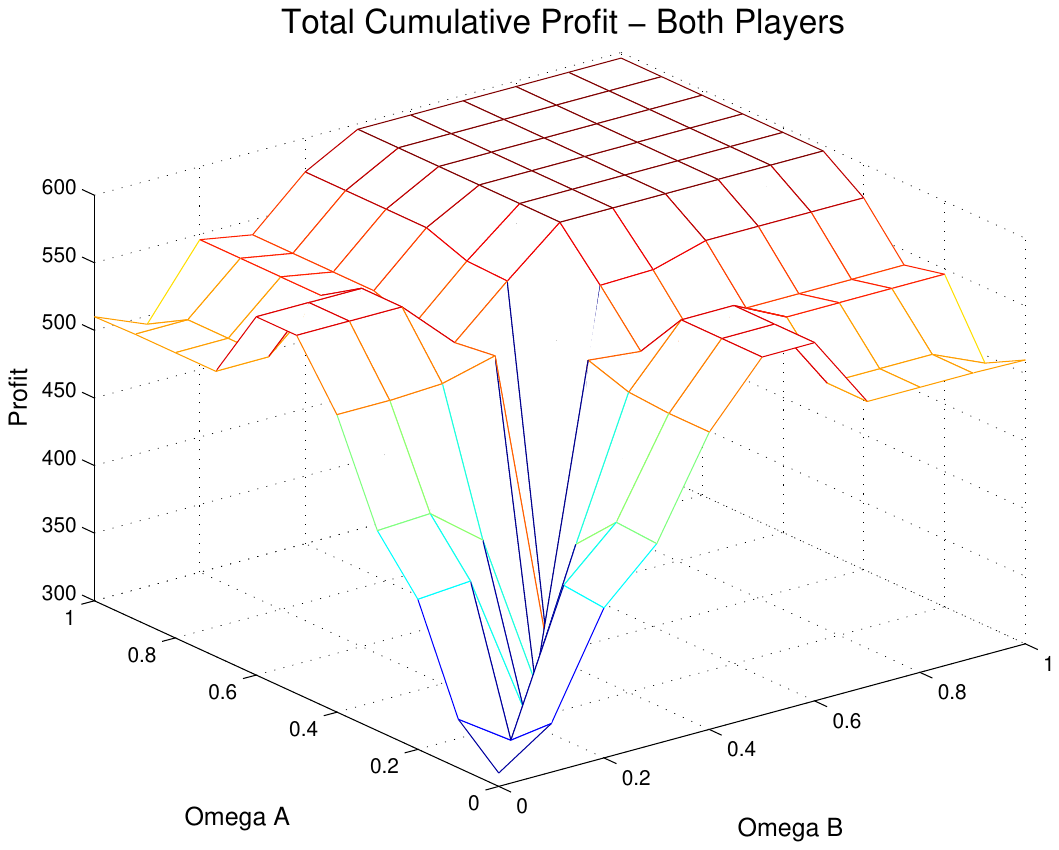}
      \caption{Test~4 - Overall profit of the players.}\label{fig:exp4_Total}
 \end{minipage}
 \hfill
  \begin{minipage}{0.49\textwidth}
    \includegraphics[trim=1.5cm 7cm 13.5cm 7cm,width=0.32\linewidth]{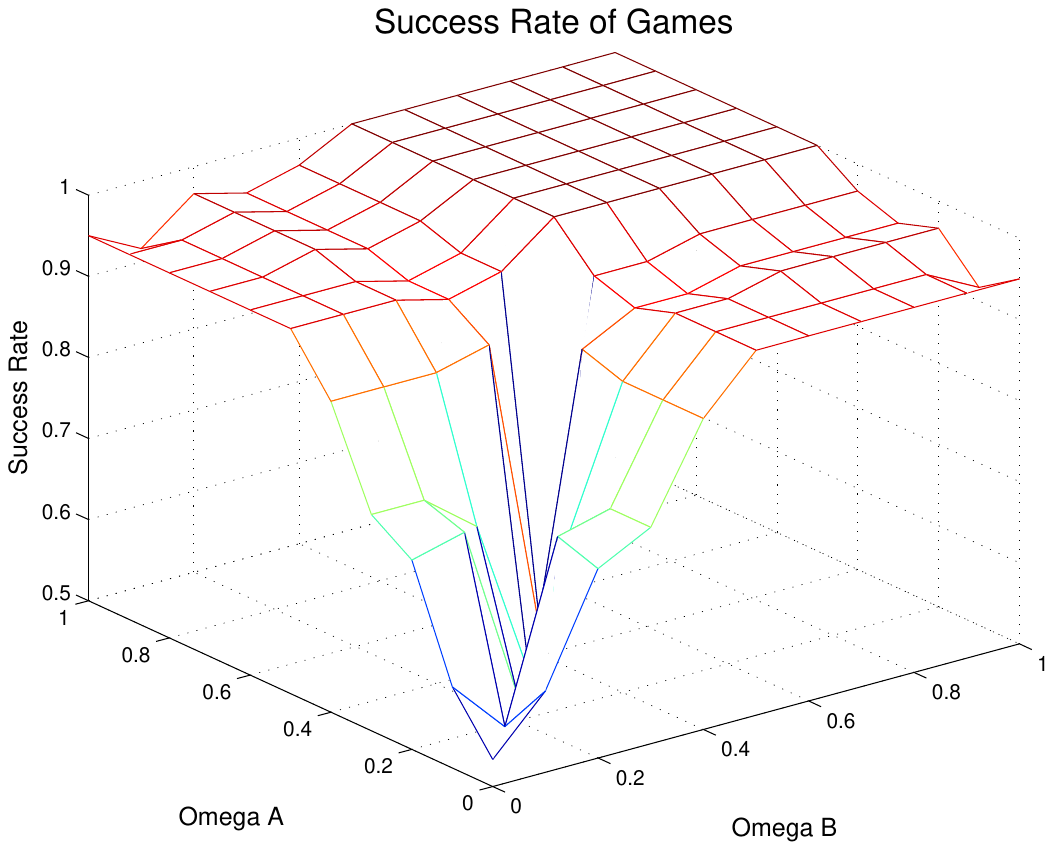}
    \caption{Test~4 - Success rate of the games.}\label{fig:exp4_Success}
 \end{minipage}
\end{figure}

Cumulative profits and success rate of the game in dependence on parameters $\omega^{\mathcal{A}}$ and $\omega^{\mathcal{B}}$ are shown in Figures~\ref{fig:exp4_Player A}-\ref{fig:exp4_Success} and Table~\ref{tab:exp5_table}.
\begin{table}
\caption{Test~4: Both players optimise and learn with uniform prior.}\label{tab:exp4_table}
\begin{center}
\begin{tabular}{|l|c|c|c|c|}
\hline
 \multicolumn{4}{|c|}{Cumulative profit} & Success rate  \\
 \hline
 & $\mathcal{A}$ & $\mathcal{B}$ & Total & \%\\
\hline
 Min. value & 152.00  & 152.00  & 304.00  & 52.00 \\
 Mean value  & 267.38  & 267.38  & 534.76  & 92.18 \\
 Max. value  & 310.00  & 310.00  & 596.00  & 100.00 \\
\hline
\end{tabular}
\end{center}
\end{table}

The results show significant improvement due to the learning. The minimum values of the individual profits and the success rate decreased but their maximum values increased on average, see Table \ref{tab:exp4_table}. The significant improvement occurred when $\omega^{\mathcal{A}}\leq 0.5$ and $\omega^{\mathcal{B}}\leq 0.5$, see Figures~\ref{fig:exp4_Player A}-\ref{fig:exp4_Total}. Compare to Test~3, learning ability brought higher individual profits as well as higher success rates achieved for relatively low values of $\omega$. Thus learning helps even when the player's willingness to negotiate (expressed by $\omega$) is low.

\subsection{Test~5: Informative prior information}\label{Exp5}
This experiment is a modification of Test~4 with each of the players having meaningful prior knowledge of the opponent. First, to get informative prior, the players played $30$ training rounds during which they gained prior models of their opponents. Then, Test~3 has been performed with the resulting prior instead of uniform distribution.

\begin{figure}[!ht]
  \begin{minipage}{0.49\textwidth}
    \includegraphics[trim=1.5cm 7cm 13.5cm 7cm,width=0.32\linewidth]{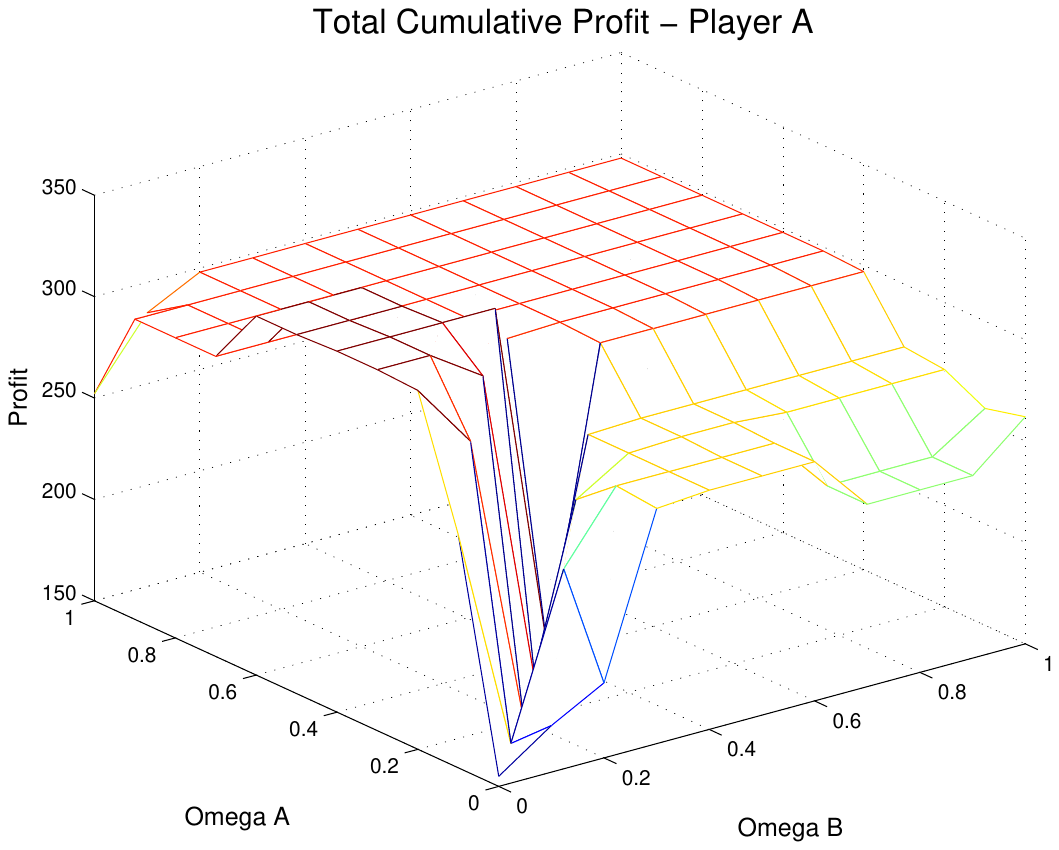}
      \caption{Test~5 -  $\mathcal{A}$'s cumulative profit in dependence on weights  $\omega^{\mathcal{A}}$ and $\omega^{\mathcal{B}}$.}\label{fig:exp5_Player A}
 \end{minipage}
 \hfill
  \begin{minipage}{0.49\textwidth}
    \includegraphics[trim=1.5cm 7cm 13.5cm 7cm,width=0.32\linewidth]{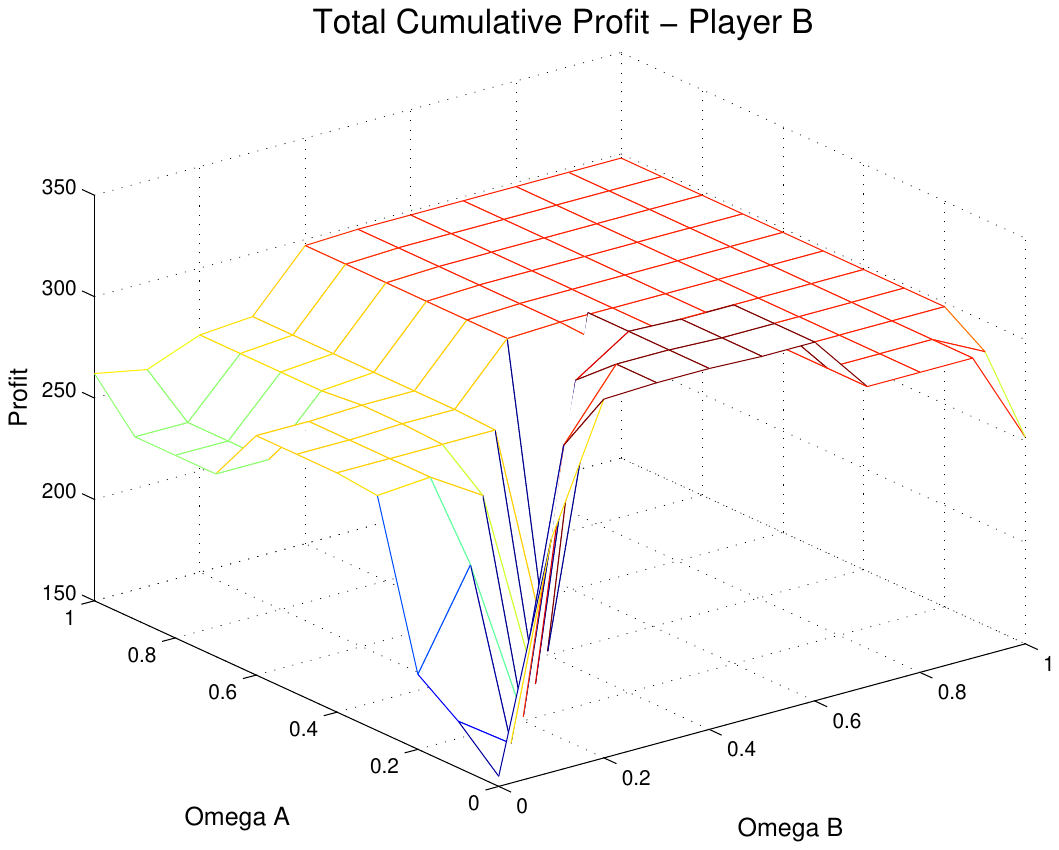}
    \caption{Test~5 -  $\mathcal{B}$'s cumulative profit in dependence on weights  $\omega^{\mathcal{A}}$ and $\omega^{\mathcal{B}}$.}\label{fig:exp5_Player B}
 \end{minipage}
\end{figure}

\begin{figure}[!ht]
  \begin{minipage}{0.49\textwidth}
    \includegraphics[trim=1.5cm 7cm 13.5cm 7cm,width=0.32\linewidth]{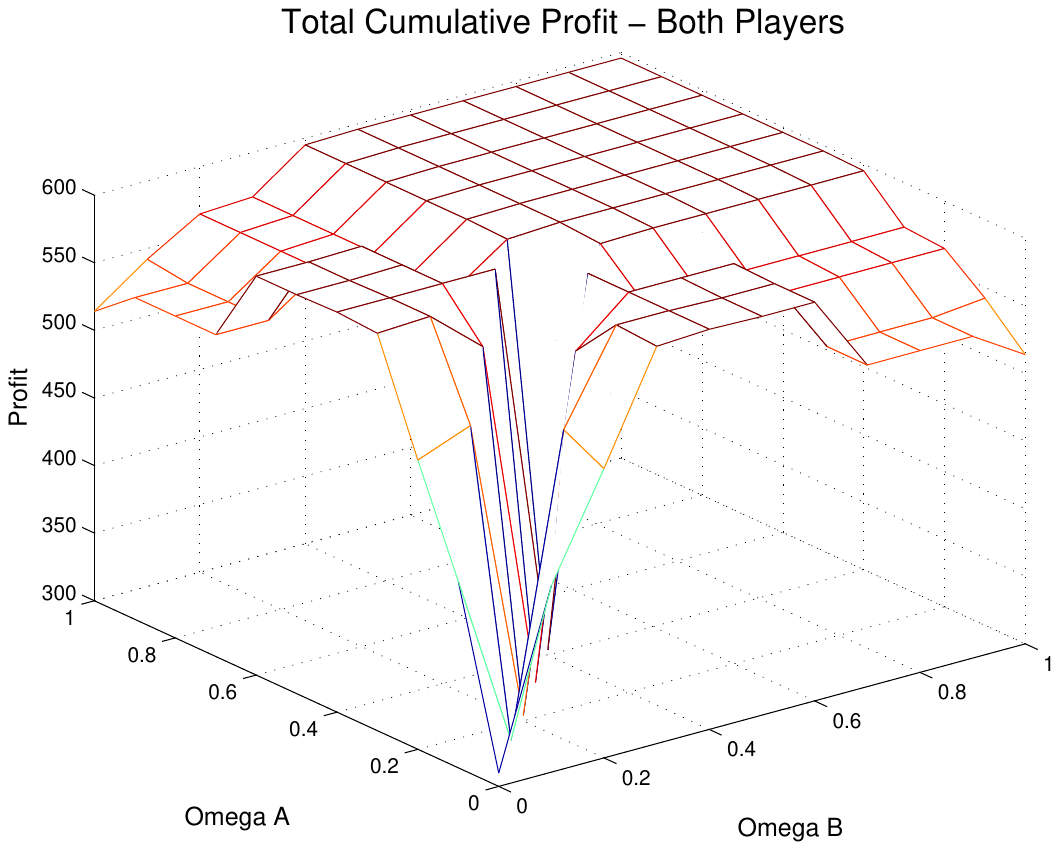}
      \caption{Test~5 - Overall profit of the players.}\label{fig:exp5_Total}
 \end{minipage}
 \hfill
  \begin{minipage}{0.49\textwidth}
    \includegraphics[trim=1.5cm 7cm 13.5cm 7cm,width=0.32\linewidth]{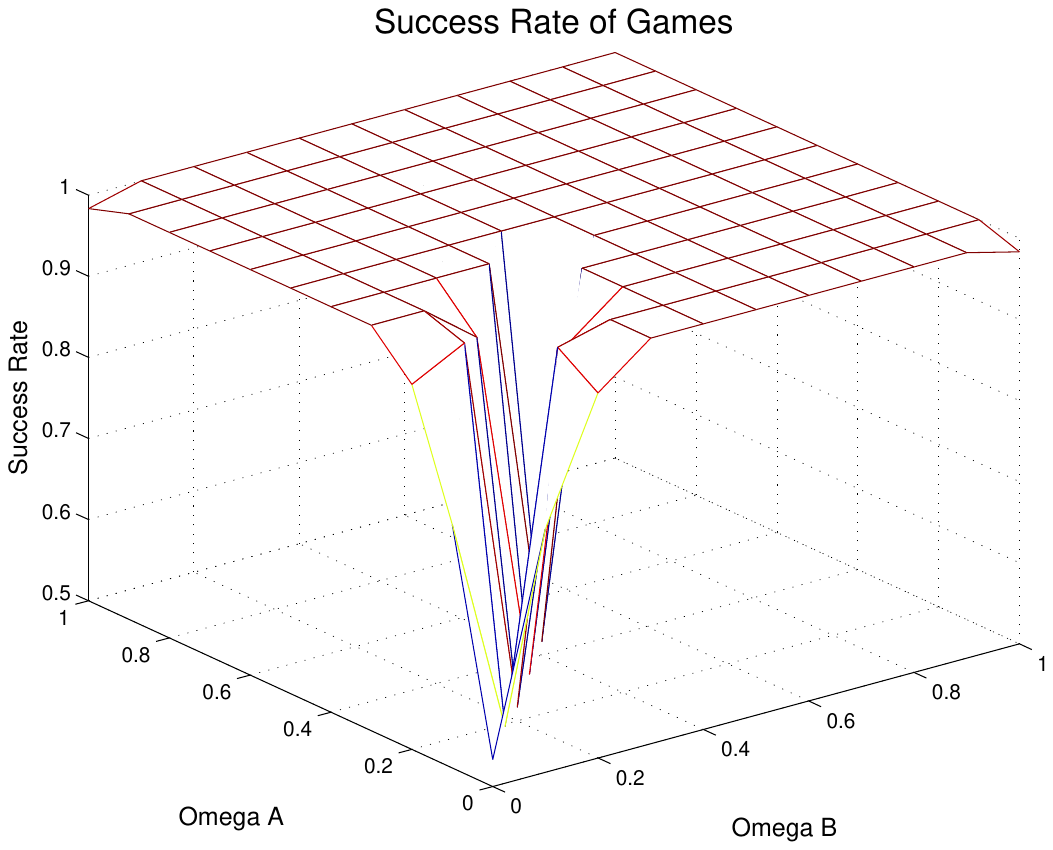}
    \caption{Test~5 - Success rate of the games.}\label{fig:exp5_Success}
 \end{minipage}
\end{figure}

Cumulative profits and success rate of the game in dependence on parameters $\omega^{\mathcal{A}}$ and $\omega^{\mathcal{B}}$ are shown in Figures~\ref{fig:exp5_Player A}-\ref{fig:exp5_Success} and Table~\ref{tab:exp5_table}.
\begin{table}
\caption{Test~5: Both players have informative prior, learn and optimise.}\label{tab:exp5_table}
\begin{center}
\begin{tabular}{|l|c|c|c|c|}
\hline
 \multicolumn{4}{|c|}{Cumulative profit} & Success rate  \\
 \hline
 & $\mathcal{A}$ & $\mathcal{B}$ & Total & \%\\
\hline
 Min. value  & 152.00  & 152.00  & 304.00  & 52.00  \\
 Mean value  & 282.40  & 282.64  & 565.27  & 97.48  \\
 Max. value  & 328.00  & 328.00  & 596.00  & 100.00\\
\hline
\end{tabular}
\end{center}
\end{table}

The results show further improvement, see Table \ref{tab:exp5_table}, cf. Tests~3-4. The minimum values of profits and success rate do not change but the maximum and mean values noticeably increased, cf. Test~4 (Section \ref{Exp4}). The players achieve much higher individual profits for low values of weights $\omega$ because they coordinated their demands to make them almost always compatible (see Figure~\ref{fig:exp5_Success}).

\section{Discussion} \label{sec:discussion}

Section~\ref{sec:Experiment} describes simulation results obtained on the NDG. It can be seen that our DM model can help the players to effectively bargain and counteract the incomplete knowledge. The main advantages of the proposed DM model are as follows:
\begin{itemize}
	\item  The proposed reward function respects individual economic profit of the bargaining agent and the unclaimed amount of money from the previous round. As the opponent's past actions enter the reward \eqref{eq:reward}, the optimal policy of the agent implicitly respects them. And vice versa: the optimal policy of the opponent respects agent's actions. Hence both players are forced to implicitly cooperate.
	
	\item The weight $\omega$ in \eqref{eq:reward} expresses trade-off between the individual profitability and efficiency of using game potential. At the same time it also reflects agent's preferences and partially style of playing (personal traits).  High values of the weight in the player's reward \eqref{eq:reward} indicate a high interest of the agent in efficient use of game resources, i.e. in minimising the remaining unclaimed amount. In each round thus the reward \textit{encourages} the agent to dynamically "adapt" its current demand to the predicted demand of the opponent. In the next round, the resulting profit\footnote{which reflects the effect of the previous round and thus provides a feedback to the agent.} together with the updated opponent's model, is used in \eqref{e:expectation}, \eqref{eq:optimal_policy} to select a new demand. This is the essence of the proposed \textit{indirect dynamic negotiation}.
	
	\item Compared to the heuristic bargaining model, Section~\ref{sec:models of B}, our optimal DM policy increased the mean value of the player's individual profit by more than $50\%$ (in the case of an uninformative prior) and by about $65\%$ (informative prior).
	
	\item Learning significantly improves the bargaining results. However optimising but not learning agent can have worse individual results compare with the heuristic opponent.
	The reason is that the optimising agent implicitly cooperates with the opponent during bargaining but does not use the correct opponent model for this\footnote{and therefore cannot predict the opponent}. On contrary, the opponent does not cooperate and it uses a fixed heuristic model. As a result, the agent's effort brings more profit to the opponent than to itself.	
	
	\item The best bargaining results were achieved if both players are learning and employ the proposed bargaining policy. Informative prior used in learning can significantly improve the agent's profit.
	\end{itemize}
The proposed solution can be further extended i) to cover multi-issue bargaining; ii) to respect human non-rationality given by social and cognitive aspects; iii) to respect emotional state of the agent that has been proved to significantly influence DM \cite{KapVolLam:17}.
\section{Concluding remarks} \label{sec:Conclusions}

The paper addresses a problem of sequential bilateral bargaining with incomplete information. We proposed DM model that helps agents to successfully bargain by performing indirect negotiation and learning the opponent's model. Methodologically the paper casts heuristically-motivated bargaining of a self-interested independent agent into a framework of Bayesian learning and Markov decision processes. The proof of the main results is based on the standard methodology. However, the problem formulation and the gained solution are novel and practically important. The special form of the reward \textit{implicitly} motivates the players to negotiate indirectly, via closed-loop interaction. At the same time the proposed method is privacy-preserving, since it does not require the exchange of data or models between the bargaining agents. We illustrate the approach by applying our model to the Nash demand game, which is an abstract model of bargaining.
The paper provides our original formulation and solution of the practically important DM scenario. It presents the initial study that confirms that our formulation is meaningful and gives the promising results.
The results indicate that the introduced DM model: i) leads to coordinating the players' actions and to their indirect negotiation; ii) results in maximising success rate of the game and iii) brings more individual profit to the players compare to the heuristic model.

The proposed bargaining policy minimises losses caused by: (i) insufficient use of the resources; (ii) demands that exceed the total resources available; and (iii) incomplete knowledge.

The results obtained indicate possibility to create a realistic and applicable methodology of cooperation and negotiation in \textit{flatly} organised networks of interacting agents without a fixed structure, cf.~\cite{WuLiuQuiHer:22}. We believe that our approach is suitable for non-cooperative, multi-agent
networks, since we provide an easy way to implicit cooperation. The solution does not rely on a central authority and the proposed DM model outperforms a heuristic model whenever both agents are rational, learning and follow the optimal strategy.

 In future work we would like :
 \begin{itemize}
 	\item to cover the multi-issue bargaining;
 	\item to extend the approach to a multi-agent settings;
 	\item to implement the approach for other bargaining rules than NDG.
 \end{itemize}
Further foreseen challenge is learning weights of individual players based on their bargaining history. The weights indirectly reflect agent's model of bargaining and preferences. Moreover the weights may depend on the agent's personality \cite{FioLinMesVil:2013}, which allows taking into account the influence of personality traits on decision making.
\section*{Acknowledgement} The authors would like to thank Miroslav K\'{a}rn\'{y} for his thorough feedback on an earlier draft and for useful discussions.


\EOD
\end{document}